\documentclass[journal]{IEEEtran}

\usepackage{cite}
\usepackage{amsthm,amssymb,amsmath}
\usepackage{mathtools}
\usepackage[final]{graphicx}

\usepackage[hidelinks]{hyperref}
\usepackage{color}
\usepackage{diagbox}

\usepackage{balance}

\DeclareMathOperator*{\argmax}{\arg\!\max}

\usepackage{algpseudocode}
\usepackage{algorithm}
\usepackage{algorithmicx}

\begin{document}
\title{Frequency-Shift Chirp Spread Spectrum Communications with Index Modulation}
\author{Muhammad Hanif,~\IEEEmembership{Senior Member,~IEEE}, and
        Ha H. Nguyen,~\IEEEmembership{Senior Member,~IEEE}%
\thanks{M. Hanif is with the Department of Engineering and Applied Science, Thompson Rivers University, Kamloops, BC V2C 0C8 (email: \texttt{mhanif@tru.ca}), and H. H. Nguyen is with the Department
of Electrical and Computer Engineering, University of Saskatchewan, Saskatoon, SK S7N 5A9,
Canada, (e-mail: \texttt{ha.nguyen@usask.ca}).}
}

\maketitle

\begin{abstract}
This paper introduces a novel frequency-shift chirp spread spectrum (FSCSS) system with index modulation (IM). By using combinations of orthogonal chirp signals for message representation, the proposed FSCSS-IM system is very flexible to design and can achieve much higher data rates than the conventional FSCSS system under the same bandwidth. The paper presents optimal detection algorithms, both coherently and non-coherently, for the proposed FSCSS-IM system. Furthermore, a low-complexity non-coherent detection algorithm is also developed to reduce the computational complexity of the receiver, which is shown to achieve near-optimal performance. Results are presented to demonstrate that the proposed system, while enabling much higher data rates, enjoys similar bit-error performance as that of the conventional FSCSS system.
\end{abstract}


\IEEEpeerreviewmaketitle

\newtheorem{theorem}{Theorem}
\newtheorem{lemma}{Lemma}
\newtheorem{definition}{Definition}
\newtheorem{corollary}{Corollary}
\newtheorem{remark}{Remark}

\section{Introduction}

\IEEEPARstart{T}{he} modulation scheme in a  chirp spread spectrum (CSS) communication system uses linear frequency-modulated chirps to represent message symbols. Due to its robustness against narrow-band interference, constant envelope and resistance against multi-path fading and Doppler effect, CSS modulation has been adopted in various low-power wide-area (i.e., long-range) wireless applications \cite{IEEE802154a,LPWANTutorial,SemtechAN22,LoRaDesignGuide,FreqShiftChirpModLoRa}. For example, the IEEE 801.15.4a standard \cite{IEEE802154a} has provisions for CSS modulation along with the burst-position modulation scheme. More recently, LoRa technology has been introduced for Internet-of-Things (IoT) applications, which is also based on frequency-shift chirp modulation and more commonly referred to as LoRa modulation \cite{LPWANTutorial,SemtechAN22,LoRaDesignGuide,FreqShiftChirpModLoRa}.

Historically, linear chirps have been used in continuous-wave frequency-modulated radars due to their good autocorrelation properties \cite{ChirpRadar}. In digital communications, CSS was first introduced in a simple form of binary modulation, in which two linear-frequency modulated chirps with opposite sweep directions, i.e., up chirp and down chirp, are used to represent an information bit. Such a scheme is also known as slope-shift keying \cite{FMChirpCommun,ChirpModDigitalSignal}. Although the up and down chirps are nearly orthogonal to each other, they are not completely orthogonal. As a consequence, the lack of complete orthogonality affects the  bit-error-rate (BER) performance of slope-shift keying.

For non-binary transmission, only one linear chirp is used in \cite{DQPSKChirp}, but its initial phase is modified using differential quadrature phase-shift keying (DQPSK) to encode the information bits. Similarly, the authors in \cite{QAMChirp} employ quadrature amplitude modulation (QAM) instead of DQPSK for encoding the information bits in both the initial phase and amplitude of a linear chirp. However, these techniques come at the expense of increased receiver complexity and signalling overhead since the phase and/or amplitude of the received chirp needs to be estimated.

One approach to reduce the receiver complexity is to embed the information in the initial frequency of the linear chirp instead of its initial phase \cite{LPWANTutorial,LoRaDesignGuide,FreqShiftChirpModLoRa}. Such a modulation scheme is aptly named as frequency-shift chirp modulation (FSCSS modulation) \cite{FreqShiftChirpModLoRa} and is also known simply as CSS when used in the context of LoRa \cite{LPWANTutorial,LoRaDesignGuide,CSSLongRange}. 

FSCSS modulation enjoys low-complexity demodulation thanks to the fast Fourier transform (FFT) \cite{FreqShiftChirpModLoRa}. However, due to a limited number of orthogonal frequency-shift chirps available in given bandwidth and symbol duration, FSCSS modulation suffers from a low data rate, especially when the available bandwidth is limited \cite{InterleavedCSS}. As such, different standards that employ CSS modulation have provisions for other modulation schemes to attain a higher data rate than what is achievable with the CSS modulation. For example, LoRa chips have two modulation schemes implemented in them: the LoRa (FSCSS) and the frequency-shift keying (FSK) modulation schemes. The data rate provided with LoRa modulation depends primarily on two factors, the spreading factor (SF) and the occupied bandwidth. The highest data rate is $37.5$ kbps when the smallest value of SF, that is ${\rm SF}=7$, and the maximum bandwidth of $500$ kHz are selected\footnote{In general, with LoRa modulation, the higher the data rate is, the lower the receiver sensitivity, i.e., the shorter the transmission range to achieve a given level of bit error rate.}. For applications that require higher data rates, LoRa chips have an FSK transceiver, but it can achieve data rates up to $50$ kbps only \cite{LoRaDesignGuide}.

Although implementing multiple transceivers in a device can increase the achievable data rate, it consumes more hardware and software resources, especially at the reception side. As such, there have been active research efforts to increase the data rate of the conventional frequency-shift CSS communication systems. For example, the authors in \cite{PSKLoRa,EfficientDesignCSS} have proposed to increase the data rate by embedding additional bits in the initial phase of an up chirp. Unfortunately, carrying additional information in the phase requires channel estimation at the receiver side, which increases the receiver complexity, not to mention the sensitivity of these schemes to synchronization errors \cite{SSKLoRa}.

In \cite{InterleavedCSS}, the authors propose to use interleaved chirps along with linear up chirps to double the number of chirp signals, hence increasing the data rate by 1 bit per each symbol (message). However, similar to binary slope-shift keying, the BER performance of the interleaved FSCSS modulation is degraded due to the non-orthogonality between the added interleaved chirps and the original set of orthogonal up chirps.

Another approach is recently introduced in \cite{SSKLoRa} in which down chirps are used instead of the interleaved chirps to double the size of the signal set. It was demonstrated in \cite{SSKLoRa}, via correlation and error-rate analysis, that using down chirps instead of interleaved chirps reduces the peak correlation between the added signal set and the original signal set, hence improving the BER performance. Nevertheless, by only doubling the original signal set, the data rate improvement attained by the two schemes in \cite{InterleavedCSS,SSKLoRa} is quite marginal, namely only one extra bit per modulated symbol. Similar to \cite{SSKLoRa}, both up and down chirps are used for data transmission in \cite{FlipLoRa}. However, instead of increasing the data rate as in \cite{SSKLoRa}, the authors in \cite{FlipLoRa} leverage the near orthogonality of the up and down chirps to improve the performance of LoRa communications under packet collisions.

In this paper, by making use of the concept of permutation modulation (PM) introduced in \cite{PermutationModulation}, we develop a different approach to increase the data rate of the conventional FSCSS modulation scheme. Specifically, an important class of permutation modulation that uses binary permutation vectors, known as index modulation (IM) \cite{IMfor5GBook}, is applied to the set of orthogonal up chirps.

It is pointed out that IM has been employed in conventional frequency-shift keying (FSK) modulation \cite{FSKPermutationModulation}, as well as in orthogonal frequency-division multiplexing (OFDM) systems \cite{OFDMIMBasar,EnhancedOFDMwithIM,MultipleModeOFDMwithIM,NoncoherentOFDMIM,GeneralizedOFDMIM}. The technique of IM appears in other names in other applications, such as parallel-combinatory (PC) spread-spectrum modulation \cite{PCSpreadSpectrum}, spatial modulation (SM) for multi-antenna systems \cite{SpatialModulationConf,SpatialModulation,GeneralizedSpatialMod,GeneralisedSpatialModMultipleAntenna}, space-shift keying \cite{SpaceShifKeyingMIMO,GeneralizedSpaceShiftKeying} multi-tone FSK \cite{MultiToneFSK}, etc. The interested reader is referred to \cite{PMSMandIMTutorial,SurveySpatialMod,IMTechniquesNextGen,MultidimensionalIMfor5GandBeyond} for a comprehensive review of index modulation and its applications.

To the best of our knowledge, this paper is the first work on the application of IM in chirp spread spectrum communications\footnote{The main idea and inventions in this paper are protected by a US patent \#10,778,282 \cite{USpatent}.}. A somewhat related technique to increase the data rate is suggested in \cite{QAMChirp,OverlapCSS} by transmitting multiple chirps together. In particular, time-delayed basic up chirps are added to the basic up chirp, and the information bits are represented in the phase and/or amplitude of the basic chirp and its time-delayed versions, while no information is represented in the initial delay of the added time-shifted basic chirp. Such a technique is markedly different from the index modulation technique in which the information is embedded in the indices of the added signals. As shall be seen, for the FSCSS modulation, those indices are related to the initial delays or the initial frequencies of the up chirps. Other techniques, such as, code index modulation or OFDM-IM spread spectrum and their variants \cite{CodeIndexModulation,GeneralizedCodeIndexModulation,OFDMIMSpreadSpectrum} apply the IM principle to select spreading codes to represent a part of information bits. These techniques are fundamentally different from the proposed CSS modulation technique, which uses chirps to represent the information bits instead of spreading codes.

The rest of the paper is organized as follows. Section \ref{sec:sys} reviews the conventional FSCSS system. Our proposed FSCSS with index modulation (FSCSS-IM) scheme is detailed in Section \ref{sec:ProposedScheme}. The optimal and suboptimal detection algorithms for the proposed FSCSS-IM scheme are developed in Section \ref{sec:Rx}. Performance results and comparison to the conventional FSCSS system are presented in Section \ref{sec:simulation}. Section \ref{sec:conclu} concludes the paper.

\section{Conventional FSCSS System} \label{sec:sys}

In a conventional FSCSS system, a set of linearly-independent up chirps are used for representing information bits. In particular, a set of $M$ orthogonal chirps are used, which are all generated from one linear up chirp, also referred to as the basic chirp in the sequel. The complex baseband-equivalent form of the basic chirp has the following discrete-time representation \cite{LoRaModIoT,EfficientDesignCSS}:
\begin{IEEEeqnarray}{c}
x_0[n] = \exp\left\{j  \frac{\pi n^2}{M}\right\},\IEEEeqnarraynumspace
\end{IEEEeqnarray}
where $n=0,1,\cdots,M-1$, and $M$ is the number of orthogonal chirps used in the conventional FSCSS modulation scheme. The other orthogonal chirps are related to $x_0[n]$ as \cite{EfficientDesignCSS}
\begin{IEEEeqnarray}{c}
x_m[n] = x_{0}[n+m],\IEEEeqnarraynumspace
\end{IEEEeqnarray}
where $0 \le n, m <M$. Observe that
\begin{IEEEeqnarray}{rl}\label{Eq:mChirpExp}
x_m[n] & = \exp\left\{j  \frac{\pi (n^2+m^2+2nm)}{M}\right\} \IEEEnonumber \\
& =  x_0[m] x_0[n] \exp\left(j \frac{2\pi m}{M}n\right).\IEEEeqnarraynumspace
\end{IEEEeqnarray}
As such, the instantaneous (digital) frequency of the $m$th chirp differs from that of the basic chirp by $\frac{m}{M}$.

In the conventional FSCSS modulation, each chirp is associated with a message symbol. Specifically, let $b_{0},b_{1},\cdots,b_{\Lambda-1}$ be a group of $\Lambda$ information bits associated with a message symbol, and $m$ be the corresponding decimal number, i.e., $m = \sum_{i=0}^{\Lambda-1} b_i 2^i$.  The transmitted signal corresponding to message $m$, denoted by $s_m[n]$, is simply given by  $s_m[n] = x_m[n]$. Note that $\Lambda  = \log_2 M$, or $M=2^{\Lambda}$. This means that in a conventional FSCSS modulation, such as in LoRa modulation, the number of chirps is selected to be a power of $2$.

Assuming perfect synchronization, the received signal, denoted by $y[n]$, is given as
\begin{IEEEeqnarray}{c}
y[n] = h x_m[n] + w[n],\IEEEeqnarraynumspace
\label{rx-yn}
\end{IEEEeqnarray}
where $m$ is an arbitrary transmitted symbol, $h$ denotes the channel gain, and $w[n]$ is a zero-mean circularly-symmetric white Gaussian noise with variance $N_0$. Since $\sum_{n=0}^{M-1}|x_m[n]|^2 = M$, the average signal-to-noise ratio (SNR) of the received symbol, denoted by $E_s/N_0$, is given by
\begin{IEEEeqnarray}{c}
E_s/N_0 = \frac{M E\left[|h|^2\right]}{N_0} = M \bar{\gamma},\IEEEeqnarraynumspace
\end{IEEEeqnarray}
where $E\left[|h|^2\right]$ is the expected value of the squared absolute value of the channel gain, and $\bar{\gamma} = \frac{E\left[|h|^2\right]}{N_0}$ is the average \emph{channel} SNR. For an additive white Gaussian noise (AWGN) channel, the channel gain is $h = 1$, and the instantaneous and average channel SNRs are both equal to $1/N_0$.

Detection of the transmitted symbol $m$ works as follows \cite{EfficientDesignCSS,FreqShiftChirpModLoRa}. First the received signal in \eqref{rx-yn} is multiplied with the complex conjugate of the basic chirp to produce $r[n]=y[n]x_0^*[n]$. Then the discrete-Fourier transform (DFT) is applied on $r[n]$ to obtain $R[l]=\sum_{n=0}^{M-1}r[n] e^{-j \frac{2\pi l}{M}n}$, $l=0,1,\cdots,M-1$.

For coherent detection, i.e., when the channel gain $h$ is available at the receiver, the ML estimate of the transmitted symbol can be found by
\begin{IEEEeqnarray}{c}\label{Eq:CoherentReceiverConvLoRa}
\hat{m} = \argmax_{m} \Re \big\{h^* x_0^*[m] R[m]\big\}.\IEEEeqnarraynumspace
\end{IEEEeqnarray}
For an AWGN channel, \eqref{Eq:CoherentReceiverConvLoRa} further simplifies to
\begin{IEEEeqnarray}{c}\label{Eq:AWGNReceiverConvLoRa}
\hat{m} = \argmax_{m} \Re \big\{x_0^*[m] R[m]\big\}.\IEEEeqnarraynumspace
\end{IEEEeqnarray}

When $h$ is not available at the receiver, the optimal non-coherent detection of FSCSS signals amounts to simply identifying the index of the peak absolute value of the DFT output. That is,
\begin{IEEEeqnarray}{c}\label{Eq:NonCoherentReceiverConvLoRa}
\hat{m} = \argmax_{m} \big|R[m]\big|^2.\IEEEeqnarraynumspace
\end{IEEEeqnarray}

\section{Proposed FSCSS with Index Modulation} \label{sec:ProposedScheme}

The proposed scheme exploits index modulation to increase the data rate of the conventional FSCSS system. In particular, instead of transmitting only one chirp at a time, multiple chirps are transmitted \emph{simultaneously}. This results in a significant expansion of the signal set, which allows embedding more bits to one transmitted symbol, thus increasing the achievable data rate. However, increasing the size of the signal set results in interference among transmitted symbols as the orthogonality among the transmitted symbols is not preserved in the enlarged signal space. This, in turn, results in deterioration of the BER performance of the resultant scheme. Additionally, the receiver's complexity generally increases since the receiver has to estimate the transmitted symbol from a larger signal set.

\begin{figure*}[!hbt]
    \centering
    \includegraphics[width=1.5\columnwidth]{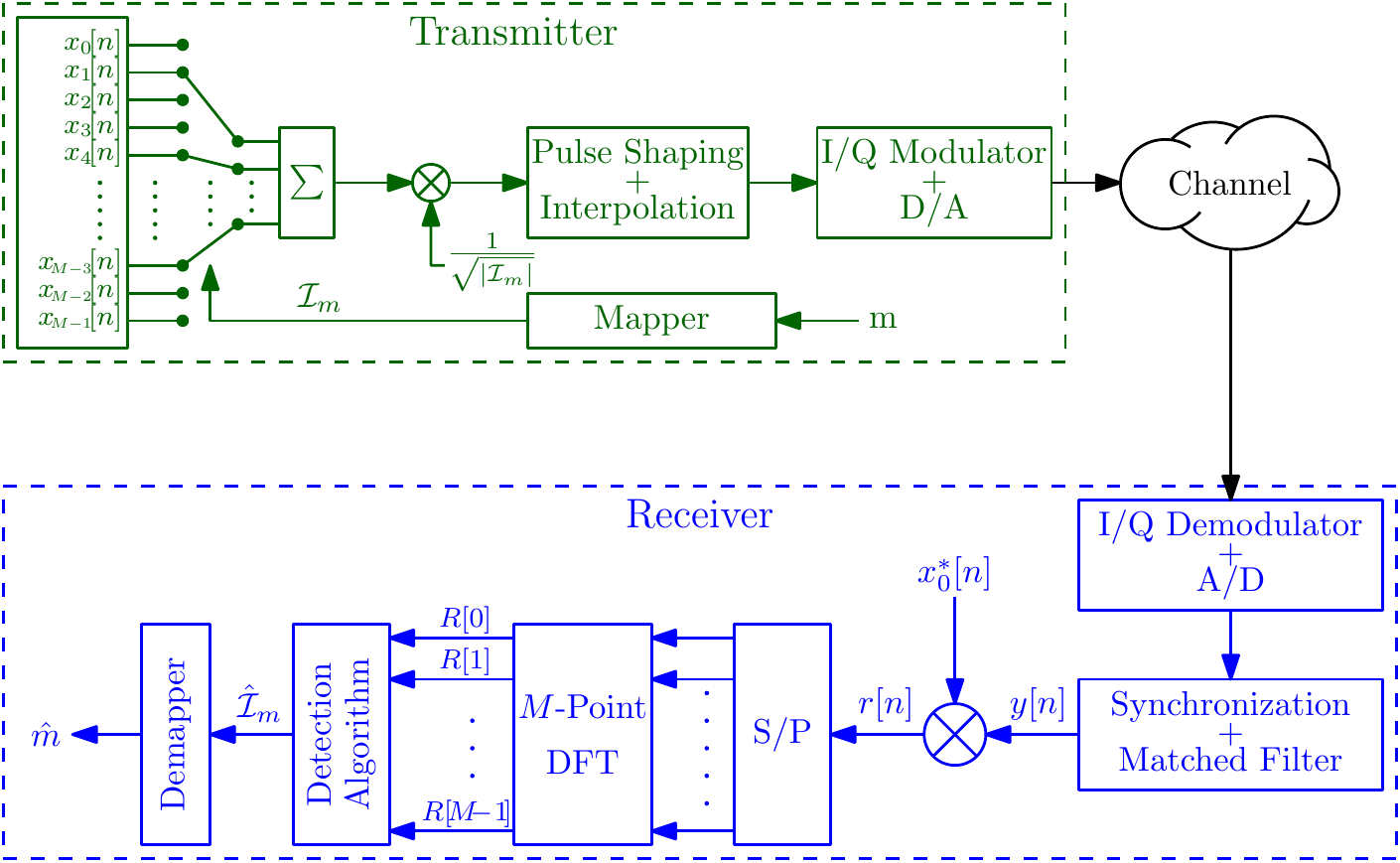}	
    \caption{Block diagram of the proposed FSCSS-IM scheme.}
    \label{Fig:BlockDiagramPropSch}
\end{figure*}

Fig. \ref{Fig:BlockDiagramPropSch} depicts the block diagram of the transmitter and receiver for the proposed FSCSS-IM system. Among the depicted blocks, the pulse shaping, interpolation, inphase-quadrature modulation and digital-to-analog conversion, are present in the transmitter of a typical communication system. Likewise, the receiver of a typical communication system has an analog-to-digital converter, inphase-quadrature demodulator, synchronizer and a matched filter. The rest of the blocks are specific to the proposed scheme. In this section we describe how the proposed scheme is implemented in the transmitter and also analyze the peak-to-average-power ratio (PAPR) of the resultant transmitted signal. The optimal and low-complexity detection algorithms for the proposed scheme are developed in Section \ref{sec:Rx}.

In the proposed scheme, the transmitter assigns a unique and distinct combination of orthogonal chirps to every value of $m$ between $0$ and $2^{\Lambda}-1$. Let $\mathcal{I}_m$ denote the index set of the chirps corresponding to message $m$. Then the transmitted signal corresponding to message $m$ is constructed as
\begin{IEEEeqnarray}{c}
\label{Eq:ProposedTransmitSignal}
s_m[n] = \frac{1}{\sqrt{|\mathcal{I}_m|}}\sum_{l \in \mathcal{I}_m}{x_{l}[n]}.\IEEEeqnarraynumspace
\end{IEEEeqnarray}
As a simple example, consider $M=8$ and $K=2$. This means that a total of $\binom{8}{2}=28$ different messages can be sent using the following combinations for index sets $\mathcal{I}_m$: $\{0, 1\}$, $\{0, 2\}$, $\cdots$, $\{0, 7\}$, $\{1, 2\}$, $\{1, 3\}$, $\cdots$, $\{1, 7\}$, $\{2, 3\}$, $\{2, 4\}$, $\cdots$, $\{2, 7\}$, $\{3, 4\}$, $\{3, 5\}$, $\cdots$, $\{3, 7\}$, $\{4, 5\}$, $\{4, 6\}$, $\{4, 7\}$, $\{5, 6\}$, $\{5, 7\}$, $\{6, 7\}$. Then the message $m=0$ corresponds to $\mathcal{I}_0= \{0, 1\}$ and can be sent using the signal $s_0[n]=\frac{1}{\sqrt{2}}(x_0[n]+x_1[n])$, whereas massage $m=15$ corresponds to $\mathcal{I}_{15}= \{2, 5\}$ and can be sent using the signal $s_{15}[n]=\frac{1}{\sqrt{2}}(x_2[n]+x_5[n])$, and so on.

Since the total number of distinct combinations of $M$ chirps used for transmission can far exceed $M$, the data rate of the proposed scheme can be significantly higher than that of the conventional FSCSS modulation scheme. To be specific, if all combinations of $K$ chirps out of $M$ chirps are used, where\footnote{We keep $K \le M/2$ since $\binom{M}{K} = \binom{M}{M-K}$, hence selecting $K > M/2$ does not result in increased data rate.} $1 \le K \le M/2$, the number of bits that are mapped to (i.e., carried by) one symbol is given as $\Lambda = \left\lfloor \log_2\binom{M}{K} \right\rfloor$. As an example, for $M=128$ and $K=2$, we have $\Lambda = 12$ bits. In contrast, with the conventional FSCSS modulation (which corresponds to $K=1$), $\Lambda=7$ bits only. This means that the data rate for $\{M=128,\;K=2\}$ is almost 1.7 times higher than what is achieved with the conventional FSCSS modulation using $M=128$ orthogonal chirps.

Table \ref{Table:DataRateImprovements} tabulates the number of bits that a symbol can carry in the proposed scheme for different values of $M$ when a fixed number of chirps are selected for transmission. Here, we have selected values of $M$ that are typically used in LoRa modulation \cite{SemtechAN26,SemtechSX127x}. Note that, in the conventional FSCSS modulation scheme, $|\mathcal{I}_m|=1$, and as such, a total of $\log_2 M$ bits are transmitted per symbol. Table \ref{Table:DataRateImprovements} also tabulates the data rate improvements by the proposed scheme. For example, with $M=2^{7}$ and $|\mathcal{I}_m|=4$, one symbol can convey a total of $23$ bits instead of only $7$ bits as in the conventional FSCSS scheme. These numbers translate to $3.29$ times higher data rate achieved by the proposed FSCSS-IM scheme. Observe that, for LoRa chips, this gain implies that the maximum achievable data rate can be $123$ kbps instead of $37.5$ kbps, which is also significantly higher than $50$ kbps achievable by the FSK transceiver implemented in LoRa chips. The table also lists the very modest data rate improvements by the schemes in \cite{InterleavedCSS} (abbreviated as ICS-LoRa) and  \cite{SSKLoRa} (abbreviated as SSK-LoRa).

\begin{table*}[!htb]
    \renewcommand{\arraystretch}{1.3}
    \centering
    \caption{Number of bits that can be transmitted in a symbol of different modulation schemes and data-rate improvement in percentage.}\label{Table:DataRateImprovements}
    \begin{tabular}{|l|c|c|c|c|c|c|c|c|c|}
    \cline{2-10}

     \multicolumn{1}{c|}{} & \textbf{FSCSS} & \multicolumn{2}{c|}{\textbf{ICS-LoRa \cite{InterleavedCSS}}} & \multicolumn{6}{c|}{\textbf{Proposed FSCSS-IM}} \\
    \cline{5-10}

     \multicolumn{1}{c|}{} & $|\mathcal{I}_m|={1}$ & \multicolumn{2}{c|}{\textbf{SSK-LoRa \cite{SSKLoRa}}}  & \multicolumn{2}{c|}{$|\mathcal{I}_m|=2$} & \multicolumn{2}{c|}{$|\mathcal{I}_m|=3$} & \multicolumn{2}{c|}{$|\mathcal{I}_m|=4$} \\
    \hline

    ${M=2^7}$ & ${7}$ & 8 & 114\% & 12 & 171\% & 18 & 257\% & 23 & 329\%  \\ \hline
    ${M=2^8}$ & ${8}$ & 9 & 112\% & 14 & 175\% & 21 & 263\% & 27 & 338\%  \\ \hline
    ${M=2^9}$ & ${9}$ & 10 & 111\% & 16 & 178\% & 24 & 267\% & 31 & 344\%  \\ \hline
    ${M=2^{10}}$& ${10}$& 11 & 110\% & 18 & 180\% & 27 & 270\% & 35 & 350\% \\ \hline
    ${M=2^{11}}$& ${11}$& 12 & 109\% & 20 & 182\% & 30 & 273\% & 39 & 355\% \\ \hline
    ${M=2^{12}}$& ${12}$& 13 & 108\% & 22 & 183\% & 33 & 275\% & 43 & 358\% \\ \hline
\end{tabular}
\end{table*}

Before closing this section, we provide the worst-case PAPR analysis of the transmitted signals produced by the proposed scheme. As noted, the significant data-rate improvements of the proposed scheme (as presented in Table \ref{Table:DataRateImprovements} and discussed earlier) result from adding multiple chirps together. Adding chirps together, however, does not preserve their constant envelope, which increases the PAPR.

In the following, we compare the PAPR of the proposed scheme with that of the conventional FSCSS scheme. To this end, the PAPR defined for the transmitted signal in complex baseband-equivalent form is
\begin{IEEEeqnarray}{c}\label{PAPR-def}
\mathrm{PAPR}_m = \frac{\max_{n}{|s_m[n]|^2}}{\sum_{n=0}^{M-1}|s_m[n]|^2/M} = \max_{n}{|s_m[n]|^2}, \IEEEeqnarraynumspace
\end{IEEEeqnarray}
where the last equality follows from \eqref{Eq:TransmittedSignalPower}. Moreover, \eqref{Eq:ProposedTransmitSignal} implies that
\begin{IEEEeqnarray}{c}
\max_{n} |s_m[n]|^2 = \frac{1}{|\mathcal{I}_m|}\max_{n} \sum_{l \in \mathcal{I}_m}\sum_{p \in \mathcal{I}_m} x_l[n] x_p[n]^*.\IEEEeqnarraynumspace
\end{IEEEeqnarray}
Using $|\sum_{i}a_i| \le \sum_{i}|a_i|$, we arrive at
\begin{IEEEeqnarray}{c}
\max_{n} |s_m[n]|^2 \le \!\frac{1}{|\mathcal{I}_m|}\max_{n}\!\!\sum_{l \in \mathcal{I}_m}\!\sum_{p \in \mathcal{I}_m}\! |x_l[n] x_p[n]^*| = |\mathcal{I}_m|.\IEEEeqnarraynumspace
\end{IEEEeqnarray}
Consequently, the PAPR of the transmitted signal in the proposed scheme can be upper bounded as
\begin{IEEEeqnarray}{c}
\mathrm{PAPR}_m \le |\mathcal{I}_m|.
\end{IEEEeqnarray}
Note that, for the conventional FSCSS systems, $|\mathcal{I}_m|=1$. In that case, the PAPR is exactly equal to its upper bound, i.e., $\mathrm{PAPR}_m = 1$. On the other hand, using a higher value of $|\mathcal{I}_m|$ achieves a larger data-rate improvement, but at the expense of a higher PAPR of the transmitted signal. However, the PAPR as defined in \eqref{PAPR-def} never exceeds $|\mathcal{I}_m|$, while  $|\mathcal{I}_m|$ is typically a small number in the proposed scheme. For example, in LTE systems, the target PAPR is between 6 dB and 8 dB. This range translates to $|\mathcal{I}_m| \le 6$, which can still result in huge data rate improvements (see Table \ref{Table:DataRateImprovements}).

\section{Detection Algorithms for the Proposed FSCSS-IM Scheme}\label{sec:Rx}

In this section we develop the detection algorithms for the proposed FSCSS-IM scheme. We will first present the optimal detection algorithms (both coherent and non-coherent), then a suboptimal non-coherent detection algorithm to reduce the computational complexity of the receiver.

Similar to the conventional FSCSS system in Section \ref{sec:sys}, the received signal in the proposed FSCSS-IM system is given as
\begin{IEEEeqnarray}{c}
y[n] = h s_m[n] + w[n],\IEEEeqnarraynumspace
\end{IEEEeqnarray}
where, as before, $m$ is an arbitrary message symbol, $h$ denotes the channel gain, and $w[n]$ is a zero-mean circularly-symmetric white Gaussian noise with variance $N_0$. Since the chirps $x_l[n]$ are orthogonal to one another, we have
\begin{IEEEeqnarray}{c}
\label{Eq:TransmittedSignalPower}
\sum_{n=0}^{M-1} |s_m[n]|^2 = \frac{1}{|\mathcal{I}_m|} \sum_{l \in \mathcal{I}_m} \sum_{n=0}^{M-1}|x_l[n]|^2 = M. \IEEEeqnarraynumspace
\end{IEEEeqnarray}
Therefore, the average SNR in the proposed system is exactly the same as that in the conventional FSCSS system.

\subsection{Optimal Coherent Receiver}
\label{sec:CoherentReceiver}
The maximum likelihood (ML) estimate of the transmitted message aims to maximize the log-likelihood function. Given the observed signal samples $\mathbf{y} = \begin{bmatrix} y[0] & y[1] & \cdots & y[M-1] \end{bmatrix}$ and channel information $h$, the likelihood function of message $m$ is given by the multi-variate complex Gaussian density function as \cite[Eq. (5.24)]{GoldsmithWirelessCommun}
\begin{IEEEeqnarray}{rl}\label{Eq:CondLikelihood}
f(m|\mathbf{y},h) &= \frac{1}{(\pi N_0)^M}\exp\left(-\frac{\sum_{n=0}^{M-1}|y[n]-h s_m[n]|^2}{N_0}\right) \IEEEnonumber \\
& = C \exp \left(\frac{2\Re\{h^* \sum_{n=0}^{M-1} y[n]s_m^*[n]\}}{N_0}\right),
\IEEEeqnarraynumspace
\end{IEEEeqnarray}
where $\Re\{\cdot\}$ denotes the real part of a complex number, and the constant $C$ is
\begin{IEEEeqnarray}{c}
C = \frac{1}{(\pi N_0)^M}\exp\left(-\frac{M|h|^2 + \sum_{n=0}^{M-1}|y[n]|^2}{N_0}\right).
\IEEEeqnarraynumspace
\end{IEEEeqnarray}
Taking logarithm of both sides and using \eqref{Eq:ProposedTransmitSignal}, the ML estimate of the message symbol, denoted by $\hat{m}$,  can be found as
\begin{IEEEeqnarray}{c}\label{Eq:MLReceiverSimple}
\hat{m} = \argmax_{m} \Re\left\{\frac{h^*}{\sqrt{|\mathcal{I}_m|}} \sum_{n=0}^{M-1}y[n]\sum_{l \in \mathcal{I}_m} x_l^*[n]\right\}.\IEEEeqnarraynumspace
\end{IEEEeqnarray}
Finally, using \eqref{Eq:mChirpExp}, we get
\begin{IEEEeqnarray}{rl}\label{Eq:MLReceiverSimpleFFT}
\hat{m} & = \argmax_{m} \Re\left\{\frac{h^*}{\sqrt{|\mathcal{I}_m|}}\sum_{l \in \mathcal{I}_m} x_0^*[l] \sum_{n=0}^{M-1}y[n] x_0^*[n] e^{-j \frac{2\pi l}{M}n}\right\} \IEEEnonumber\\
& = \argmax_{m} \Re\left\{\frac{h^*}{\sqrt{|\mathcal{I}_m|}}\sum_{l \in \mathcal{I}_m} x_0^*[l] R[l]\right\},
\IEEEeqnarraynumspace
\end{IEEEeqnarray}
where $R[l]$, $l=0,1,\cdots,M-1$, is the discrete-Fourier transform (DFT) of $r[n] = y[n]x_0^*[n]$ evaluated at the $l$th index. In other words, the optimal coherent ML receiver first correlates the received signal with a basic down chirp ($x_0^*[n]$), then performs the DFT operation on the correlated output and estimates the transmitted message by maximizing a weighted sum of the DFT values.

\begin{remark}
For the conventional FSCSS modulation, $\mathcal{I}_m = \{m\}$. As such, the ML estimate of the transmitted message simplifies to \eqref{Eq:CoherentReceiverConvLoRa}.
\end{remark}

\subsection{Optimal Non-Coherent Receiver}
\label{sec:NonCoherentReceiver}

When the receiver does not have the knowledge of $h$, which is the case of practical interest, the ML receiver estimates $m$ by maximizing the likelihood function $f(m|\mathbf{y}) = E[f(m|\mathbf{y},h)]$, where the expectation is carried over the distribution of $h$.
For a Rayleigh fading environment, $h$ is a circularly-symmetric Gaussian random variable \cite[Section 3.2]{GoldsmithWirelessCommun}. That is, $h = h_R + j h_I$, where both $h_R$ and $h_I$ are independent and identically distributed zero-mean Gaussian random variables with variance $\frac{\bar{\gamma}N_0}{2}$. In order to compute $f(m|\mathbf{y})$, we first note that
\begin{IEEEeqnarray}{rl}
& \frac{1}{\pi \sigma^2}\int_{-\infty}^{\infty} \int_{-\infty}^{\infty} \mathrm{e}^{-|h|^2 + 2\Re\{h^* \beta \}-\frac{|h|^2}{\sigma^2}} \mathrm{d} h_r \mathrm{d} h_i \IEEEnonumber \\
& =  \frac{\mathrm{e}^{\frac{\sigma^2}{1+\sigma^2}|\beta|^2}}{\pi \sigma^2}\int_{-\infty}^{\infty} \int_{-\infty}^{\infty} \!\!\mathrm{e}^{-\frac{\left|h-\frac{\sigma^2}{1+\sigma^2} \beta\right|^2}{\sigma^2/(1+\sigma^2)}} \mathrm{d} h_r \mathrm{d} h_i = \frac{\mathrm{e}^{\frac{\sigma^2}{1+\sigma^2}|\beta|^2}}{1+\sigma^2},
\IEEEeqnarraynumspace
\end{IEEEeqnarray}
where the last equality follows from the fact that the area under a probability-density function is unity.

Consequently, it is not hard to verify that $f(m|\mathbf{y})$ is given by
\begin{IEEEeqnarray}{c}
f(m|\mathbf{y}) = C' \exp\bigg( \frac{\bar{\gamma}N_0}{1+\bar{\gamma}N_0} \left|\sum_{n=0}^{M-1}y[n]s_m^*[n]\bigg|^2\right),\IEEEeqnarraynumspace
\end{IEEEeqnarray}
where the constant $C'$ is
\begin{IEEEeqnarray}{c}
C'  = \frac{\exp\left(-\frac{\sum_{n=0}^{M-1}|y[n]|^2}{N_0}\right)}{(\pi N_0)^M(1+\bar{\gamma}N_0)}.\IEEEeqnarraynumspace
\end{IEEEeqnarray}

Using \eqref{Eq:ProposedTransmitSignal}, the ML estimate of $m$ can be found as
\begin{IEEEeqnarray}{rl}\label{Eq:NonCoherentReceiver}
\hat{m} & =  \argmax_{m} \bigg|\sum_{n=0}^{M-1}y[n]s_m^*[n]\bigg|^2 \IEEEnonumber\\
& = \argmax_{m} \frac{1}{|\mathcal{I}_m|}\bigg|\sum_{l \in \mathcal{I}_m}\sum_{n=0}^{M-1}y[n]x_l^*[n]\bigg|^2.\IEEEeqnarraynumspace
\end{IEEEeqnarray}

Lastly, using  $x_l[n] = x_0[l]x_0[n]\exp{\left(j\frac{2\pi l}{M}n\right)}$, we obtain
\begin{IEEEeqnarray}{rl}\label{Eq:NonCoherentReceiverFFT}
\hat{m} & = \argmax_{m} \frac{1}{|\mathcal{I}_m|}\bigg|\sum_{l \in \mathcal{I}_m}x_0^*[l]\sum_{n=0}^{M-1}r[n]e^{-j2\pi n l/M}\bigg|^2 \IEEEnonumber \\
 & = \argmax_{m} \frac{1}{|\mathcal{I}_m|}\bigg|\sum_{l \in \mathcal{I}_m}x_0^*[l]R[l]\bigg|^2,\IEEEeqnarraynumspace
\end{IEEEeqnarray}
where $r[n] = y[n]x_0^*[n]$, and $R[l]$, $l=0,1,\cdots,M-1$, is the discrete Fourier transform (DFT) of $r[n]$.

\begin{remark}
When the index sets have the same cardinality for all messages, the optimal ML estimate of $m$ can be found as
\begin{IEEEeqnarray}{c}
\label{Eq:IndexSetSameCardinality}
\hat{m} = \argmax_{m} \bigg|\sum_{l \in \mathcal{I}_m}x_0^*[l]R[l]\bigg|^2. \IEEEeqnarraynumspace
\end{IEEEeqnarray}
\end{remark}

\begin{remark}
For the conventional FSCSS modulation, $\mathcal{I}_m~=~\{m\}$. Since $|x_0^*[m]|=1$, the detection rule in \eqref{Eq:NonCoherentReceiverFFT} becomes \eqref{Eq:NonCoherentReceiverConvLoRa}, as it should be.
\end{remark}

In the general case when more than one chirp are combined and used for signal transmission (hence increasing the data rate), the above optimal receiver would require a search over all the patterns used at the transmitter to find $\hat{m}$ that maximizes $\frac{1}{|\mathcal{I}_m|}\bigg|\sum_{l \in \mathcal{I}_m}x_0^*[l]R[l]\bigg|^2$. As such, the optimal receiver requires large computational effort and memory. In the sequel, we develop a suboptimal detection scheme that significantly reduces both the computational complexity and memory.

\subsection{Proposed Suboptimal Non-Coherent Receiver}\label{sec:PropSubOptimalRec}

The proposed low-complexity detection algorithm estimates $\mathcal{I}_m$ in a recursive manner. Once the set $\mathcal{I}_m$ is found, the message $m$ can be determined from $\mathcal{I}_m$ by using the combinatorial method described in \cite[Sect. III]{OFDMIMBasar}, i.e., without the need to store any lookup tables. In particular, the proposed algorithm estimates the elements of $\mathcal{I}_m$ one at a time. In order to conveniently illustrate the proposed detection algorithm, we first consider a case when all messages are represented by two chirps, i.e., $|\mathcal{I}_m| = 2$. Let $\mathcal{I}_m = \{l_1,l_2\}$. Then the proposed detection algorithm first estimates $l_1$ as
\begin{IEEEeqnarray}{c}\label{Eq:FirstChirp}
  \hat{l}_1 = \argmax_{l} |R[l]|^2. \IEEEeqnarraynumspace
\end{IEEEeqnarray}
Note that \eqref{Eq:FirstChirp} estimates $l_1$ exactly the same way as is done for the conventional FSCSS modulation (see \eqref{Eq:NonCoherentReceiverConvLoRa}). Once $l_1$ is estimated, $l_2$ can be estimated using \eqref{Eq:NonCoherentReceiverFFT} as
\begin{IEEEeqnarray}{c}\label{Eq:SecondChirp}
  \hat{l}_2 = \argmax_{l} \bigg|x_0^*[\hat{l}_1]R[\hat{l}_1]+x_0^*[l]R[l]\bigg|^2.\IEEEeqnarraynumspace
\end{IEEEeqnarray}

For $|\mathcal{I}_m|>2$, the proposed detection rule is carried out similarly. In particular, the elements of $\mathcal{I}_m=\{l_1,l_2,\cdots,l_{|\mathcal{I}_m|}\}$ are detected recursively as
\begin{eqnarray}
  \hat{l}_1 &=&\argmax_{l} \big|x_0^*[l]R[l]\big| = \argmax_{l} \big|x_0^*[l]\big|\big|R[l]\big|\nonumber\\
   &=&\argmax_{l} \big|R[l]\big|^2, \label{Eq:FirstElement} \IEEEeqnarraynumspace
\end{eqnarray}
and
\begin{IEEEeqnarray}{c}\label{Eq:RemElement}
  \hat{l}_{k+1} = \argmax_{l} \bigg|x_0^*[l]R[l]+\sum_{m=1}^{k}x_0^{*}[\hat{l}_m]R[\hat{l}_m]\bigg|^2,\IEEEeqnarraynumspace
\end{IEEEeqnarray}
for $k=1,2,\cdots,|\mathcal{I}_m|-1$.

A pseudo code of the proposed detection algorithm is described in Algorithm \ref{Alg:ProposedReceiverComplete}. The code assumes that \emph{all combinations} of the chirps (that is $\binom{M}{K}$ of them) are used for transmission. In practice, however, a total of $2^\Lambda$ codewords are used, where $\Lambda$ is the number of bits represented by each symbol. The most common case is when $\Lambda = \lfloor \binom{M}{K}\rfloor$. This means that Algorithm \ref{Alg:ProposedReceiverComplete} can produce catastrophic results when the estimated $\mathcal{I}_m$ does not belong to the codebook used at the transmitter \cite{OFDMIMBasar}. In order to avoid such catastrophic results, we introduce Algorithm \ref{Alg:ProposedReceiverPractical} and explain how it works and differs from Algorithm \ref{Alg:ProposedReceiverComplete} in the following.

To elaborate the difference between Algorithm \ref{Alg:ProposedReceiverComplete} and Algorithm \ref{Alg:ProposedReceiverPractical}, we consider Table \ref{Table:All2Outof8Comb}, which lists all $28$ combinations for $M=8$ and $|\mathcal{I}_m|=2$. Algorithm \ref{Alg:ProposedReceiverComplete} assumes that all combinations are used for message transmission, whereas Algorithm \ref{Alg:ProposedReceiverPractical} only considers the first $16$ codewords, corresponding to $m=0,1,\cdots,15$, for messages.

\begin{algorithm}
\caption{Proposed Suboptimal Receiver for the \emph{Complete} Codebook}
\label{Alg:ProposedReceiverComplete}
\begin{algorithmic}[1]
\Require{$|\mathcal{I}_m|$, $y[n]$ for $n=0,1,\cdots,M-1$}
\Ensure{$\hat{\mathcal{I}}_m$}
\State Compute $r[n] = y[n]x_0^*[n]$ for $n=0,1,\cdots,M-1$
\State Compute $\mathbf{R}= \mathrm{DFT}(\begin{bmatrix} r[0] & r[1] & \cdots & r[M-1]\end{bmatrix})$
\State Compute $Y[l] = x_0^*[l]R[l]$ for $l=0,1,\cdots,M-1$
\State $\mathcal{S} \gets \{0, 1,\cdots, M-1\}$
\State $\hat{l}_1 = \argmax_{l \in \mathcal{S}} \bigg|Y[l]\bigg|$
\State $\hat{\mathcal{I}}_m \gets \{ \hat{l}_1 \}$
\State $\mathcal{S} \gets \mathcal{S} \setminus \{ \hat{l}_1 \}$
\For{$k=2,\cdots,|\mathcal{I}_{m}|$}
\State $\hat{l}_k = \argmax_{l \in \mathcal{S}} \bigg|Y[l]+\sum_{m=1}^{k-1}Y[\hat{l}_m]\bigg|$
\State $\hat{\mathcal{I}}_m \gets \hat{\mathcal{I}}_m \cup \{\hat{l}_k\}$
\State $\mathcal{S} \gets \mathcal{S} \setminus \{ \hat{l}_k \}$
\EndFor
\end{algorithmic}
\end{algorithm}

Table \ref{Table:All2Outof8Comb} also tabulates a number $p$ along with the index set $\mathcal{I}_m$ corresponding to message $m$. For a given $\mathcal{I}_m = \{l_1, \cdots, l_{|\mathcal{I}_m|}\}$, where $l_1 < \cdots < l_{|\mathcal{I}_m|}$, $p$ is calculated as
\begin{IEEEeqnarray}{c}
p = \sum_{i=0}^{|\mathcal{I}_m|-1} l_{i+1} M^{|\mathcal{I}_m|-i}.
\IEEEeqnarraynumspace
\end{IEEEeqnarray}

Observe that there is a one-to-one relationship between $\mathcal{I}_m$ (or equivalently, $m$) and $p$, and $p$ increases as $m$ increases and vice versa. As such, $m \le m_0$ is equivalent to $p \le p_0$, where $p_0$ is the number corresponding to $\mathcal{I}_{m_0}$. For example, $m \le 15$ corresponds to $p \le 21$ for the codewords in Table \ref{Table:All2Outof8Comb}.

To work with the reduced codebook, Algorithm \ref{Alg:ProposedReceiverComplete} needs to incorporate the condition $m \le 2^{\Lambda}-1$, which is equivalent to incorporating the condition $p \le p_0$, where $p_0$ is computed for $\mathcal{I}_{2^{\Lambda}-1}$. To this end, we add an extra condition while determining the last index in the estimated $\mathcal{I}_m$ in the proposed detection scheme presented in Algorithm \ref{Alg:ProposedReceiverComplete}. The modification is incorporated in Algorithm \ref{Alg:ProposedReceiverPractical}, where we use a function $\textsc{FinalIndex}$ to determine the range of the last index, $\hat{l}_{|\mathcal{I}_m|}$.

\begin{table}
  \renewcommand{\arraystretch}{1.3}
  \renewcommand{\tabcolsep}{3pt}
  \centering
  \caption{$\mathcal{I}_m$ along with $m$ and $p$ corresponding to $2$ out of $8$ selected indexes.}\label{Table:All2Outof8Comb}
  \begin{tabular}{|c|c|c||c|c|c||c|c|c||c|c|c|}
    \hline
    $m$ & $\mathcal{I}_m$ & $p$ & $m$ & $\mathcal{I}_m$ & $p$ & $m$ & $\mathcal{I}_m$ & $p$ & $m$ & $\mathcal{I}_m$ & $p$ \\ \hline
    0 & \{0, 1\} & 1 & 7  & \{1, 2\} & 10 & 14 & \{2, 4\} & 20 & 21 & \{3, 7\} & 31 \\ \hline
    1 & \{0, 2\} & 2 & 8  & \{1, 3\} & 11 & 15 & \{2, 5\} & 21 & 22 & \{4, 5\} & 37 \\ \hline
    2 & \{0, 3\} & 3 & 9  & \{1, 4\} & 12 & 16 & \{2, 6\} & 22 & 23 & \{4, 6\} & 38 \\ \hline
    3 & \{0, 4\} & 4 & 10 & \{1, 5\} & 13 & 17 & \{2, 7\} & 23 & 24 & \{4, 7\} & 39 \\ \hline
    4 & \{0, 5\} & 5 & 11 & \{1, 6\} & 14 & 18 & \{3, 4\} & 28 & 25 & \{5, 6\} & 46 \\ \hline
    5 & \{0, 6\} & 6 & 12 & \{1, 7\} & 15 & 19 & \{3, 5\} & 29 & 26 & \{5, 7\} & 47 \\ \hline
    6 & \{0, 7\} & 7 & 13 & \{2, 3\} & 19 & 20 & \{3, 6\} & 30 & 27 & \{6, 7\} & 55 \\ \hline

  \end{tabular}
\end{table}

\begin{algorithm}
\caption{Proposed Suboptimal Receiver for a Reduced Codebook}
\label{Alg:ProposedReceiverPractical}
\begin{algorithmic}[1]
\Require{$|\mathcal{I}_m|$, $y[n]$ for $n=0,1,\cdots,M-1$, LastCode}
\Ensure{$\hat{\mathcal{I}}_m$}
\State Compute $r[n] = y[n]x_0^*[n]$ for $n=0,1,\cdots,M-1$
\State Compute $\mathbf{R}= \mathrm{DFT}(\begin{bmatrix} r[0] & r[1] & \cdots & r[M-1]\end{bmatrix})$
\State Compute $Y[l] = x_0^*[l]R[l]$ for $l=0,1,\cdots,M-1$
\State $\mathcal{S} \gets \{0, 1,\cdots, M-1\}$

\State $\hat{l}_1 = \argmax_{l \in \mathcal{S}} \bigg|Y[l]\bigg|$
\State $\hat{\mathcal{I}}_m \gets \{ \hat{l}_1 \}$
\State $\mathcal{S} \gets \mathcal{S} \setminus \{ \hat{l}_1 \}$
\For{$k=2,\cdots,|\mathcal{I}_{m}|-1$}
\State $\hat{l}_k = \argmax_{l \in \mathcal{S}} \bigg|Y[l]+\sum_{m=1}^{k-1}Y[\hat{l}_m]\bigg|$
\State $\hat{\mathcal{I}}_m \gets \hat{\mathcal{I}}_m \cup \{\hat{l}_k\}$
\State $\mathcal{S} \gets \mathcal{S} \setminus \{ \hat{l}_k \}$
\EndFor
\State Sort $\hat{\mathcal{I}}_m$ in ascending order
\State $z \gets$ \textsc{FinalIndex}($\hat{\mathcal{I}}_m$, LastCode, $|\mathcal{I}_m|$)
\State $\hat{l}_{|\mathcal{I}_m|} = \argmax_{l \in \mathcal{S}, l \le z} \bigg|Y[l]+\sum_{m=1}^{|\mathcal{I}_m|-1}Y[\hat{l}_m]\bigg|$
\State $\hat{\mathcal{I}}_m \gets \hat{\mathcal{I}}_m \cup \{\hat{l}_{|\mathcal{I}_m|}\}$
\end{algorithmic}
\end{algorithm}

The function $\textsc{FinalIndex}$ is given in Algorithm \ref{Alg:FinalIndex}. It takes three inputs: the estimated $\mathcal{I}$, the final index set $\mathcal{F}$, and the cardinality of $\mathcal{F}$. Elements of both sets are in ascending order. In the following, we explain the operations carried out by the function $\textsc{FinalIndex}$ by considering the codebook of Table \ref{Table:All2Outof8Comb}. For convenience, we denote the $k$th elements of $\mathcal{I}$ and $\mathcal{F}$ by $\mathcal{I}(k)$ and $\mathcal{F}(k)$, respectively.

For the codebook of Table \ref{Table:All2Outof8Comb}, $\mathcal{F} = \{2,5\}$. We consider four values of $\hat{l}_1$, i.e., $\hat{l}_1 = 0, 2, 4, 6$. Note that $\mathcal{I} = \{\hat{l}_1\}$. Now, for $\mathcal{I}(1) = \hat{l}_1 = 0 < \mathcal{F}(1)$, $\hat{l}_2$ can be as large as $M-1=7$. For $\mathcal{I}(1) = \hat{l}_1 = 2 = \mathcal{F}(1)$, the maximum value of $\hat{l}_2$ is $5$. Finally, the maximum value of $\hat{l}_2$ is $2$ when $\hat{l}_1 = 4$ and is $1$ when $\hat{l}_1 = 6$. Observe that, in the last two case, $\mathcal{I}(1) > \mathcal{F}(1)$, and the maximum value of $\hat{l}_2$ can be determined by ensuring that the number $p$ corresponding to $\mathcal{I}$ must not exceed $p_0 = 21$. This can be done by satisfying  $\hat{l}_2 \times M + \hat{l}_1 \le \mathcal{F}(1) \times M + \mathcal{F}(2)$. Solving the inequality yields
$\hat{l}_2 \le \mathcal{F}(1) + \mathcal{F}(2)\times M^{-1} - \mathcal{I}(1)\times M^{-1}$.

\begin{algorithm}
\caption{Final Index Range Finder}
\label{Alg:FinalIndex}
\begin{algorithmic}[1]
\Function{$\textsc{FinalIndex}$}{$\mathcal{I}$, $\mathcal{F}$, $K$}
    \If {$K = 1$}
        \State \Return{$\mathcal{F}(1)$}
    \EndIf
    \If{$\mathcal{I}(1) < \mathcal{F}(1)$}
        \State \Return{$M-1$}
    \Else
        \If{$\mathcal{I}(1) > \mathcal{F}(1)$}
            \State $p_0 \gets \mathcal{F}(1)+ \sum_{i=1}^{K-1}{\mathcal{F}(i+1)M^{-i}}$
            \State $p \gets \sum_{i=1}^{K-1}{\mathcal{I}(i)M^{-i}}$
            \State \Return{$p_0-p$}
        \Else
            \State \Return{$\textsc{FinalIndex}$($\mathcal{I}$(2:$K$), $\mathcal{F}$(2:$K$), $K$-1)}
        \EndIf
    \EndIf
\EndFunction
\end{algorithmic}
\end{algorithm}

Now, we describe the operations of $\textsc{FinalIndex}(\mathcal{I},\mathcal{F},K)$ for general $\mathcal{F}$ and $K$. Depending on the first (and the minimum) elements of both sets, we have three cases: $\mathcal{I}(1) < \mathcal{F}(1)$, $\mathcal{I}(1) > \mathcal{F}(1)$, and $\mathcal{I}(1) = \mathcal{F}(1)$. When $\mathcal{I}(1) < \mathcal{F}(1)$, we return $M-1$. On the other hand, when $\mathcal{I}(1) > \mathcal{F}(1)$, we return $\mathcal{F}(1) + \sum_{i=1}^{K-1}\mathcal{F}(i+1) M^{-i} -  \sum_{i=1}^{K-1}\mathcal{I}(i) M^{-i}$ to ensure that the number $p$ corresponding to $\mathcal{I}$ is less than or equal to $p_0$. Last, when $\mathcal{I}(1) = \mathcal{F}(1)$, finding the maximum possible value of the last index becomes equivalent to finding the maximum possible value of the last index of $\mathcal{I}(2:K)$. As such, we call $\textsc{FinalIndex}()$ to determine the maximum limit of the last index of $\mathcal{I}$ in Algorithm \ref{Alg:FinalIndex}.

\subsection{Computational Complexity and Memory Requirements}
\label{sec:ComputationComplexity}
The optimal detection algorithms determines the index set by searching over the entire codebook. Since the codebook length is of the order of $\binom{M}{|\mathcal{I}_m|} = \mathcal{O}\left(M^{|\mathcal{I}_m|}\right)$, the computational complexity and the memory requirements of the optimal algorithm are both $\mathcal{O}\left(M^{|\mathcal{I}_m|}\right)$. On the other hand, the proposed suboptimal detection algorithm does not require the codebook to be saved, which results in a huge memory saving. Furthermore, the proposed suboptimal detection algorithm determines the index set by finding the maximum of a metric for a total of $|\mathcal{I}_m|$ times. Since the complexity of finding the maximum from a list of size $M$ is  $\mathcal{O}(M)$, the computational complexity of the proposed suboptimal detection algorithm is only $\mathcal{O}\left(M |\mathcal{I}_m|\right)$, i.e., linear in both $|\mathcal{I}_m|$ and $M$ and much lower than the computational complexity of the optimal detection algorithm.

\section{Performance Results and Comparison}
\label{sec:simulation}

In this section, we present simulation results to depict the BER performance of the proposed FSCSS-IM system under both optimal and suboptimal detection algorithms. The BER performance is also compared to that of the conventional FSCSS system.

Fig. \ref{Fig:BERSF=7,8,9,k=2OptimalVsProp} depicts the BER of the proposed FSCSS-IM system in an AWGN channel when two chirps are selected for representing a message symbol. Here, the BER performance of the optimal non-coherent detection rule in \eqref{Eq:NonCoherentReceiverFFT} and the proposed low-complexity suboptimal detection algorithm described in Section \ref{sec:PropSubOptimalRec} are plotted. A total of $10^9$ Monte-Carlo trials were used to generate the plots.

\begin{figure}[t!]
\centering
\includegraphics[width=\columnwidth]{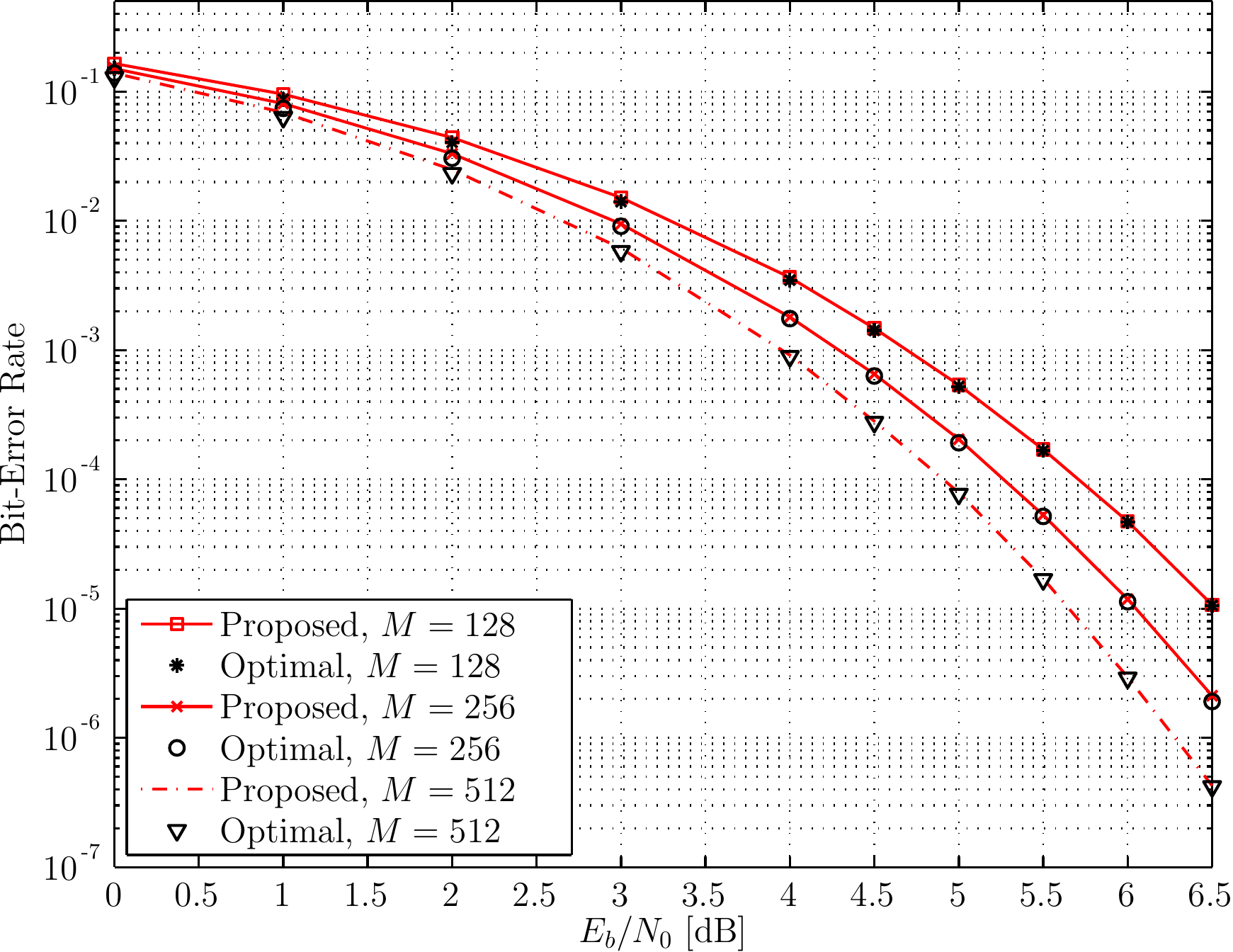}	
\caption{BER comparison of the optimal and proposed suboptimal non-coherent detection algorithms for the FSCSS-IM system when $M=128, 256, 512$, and $|\mathcal{I}_m|=2$.}
\label{Fig:BERSF=7,8,9,k=2OptimalVsProp}
\end{figure}

A couple of interesting observations can be made from the plotted curves. First, the proposed suboptimal detection algorithm enjoys near-optimal performance for the whole $E_b/N_0$ range, where $E_b/N_0 = \frac{E_s/N_0}{\Lambda}$ is the signal-to-noise ratio per bit, and for all values of $M$. This is despite the fact that the proposed scheme is highly memory efficient and significantly less computationally intensive than the optimal algorithms, as mentioned in Section \ref{sec:ComputationComplexity}.

Second, similar to the conventional FSCSS modulation, the performance of the proposed FSCSS-IM scheme improves as the number of orthogonal chirps increases. This improvement in the BER performance comes at the expense of the achievable data rate. For example, for the same occupied bandwidth, an FSCSS-IM symbol for $M=256$ is two times longer than a FSCSS-IM symbol corresponding to $M=128$. Consequently, for $|\mathcal{I}_m| = 2$, the maximum achievable data rate for $M=256$ is $1/2 \times 14/12 \approx 0.583$ times the maximum data rate achievable for $M=128$. But, on the other hand, FSCSS-IM with $M=256$ shows a performance gain of $0.5$ dB over FSCSS-IM with $M=128$ at the BER of $10^{-5}$.


Fig. \ref{Fig:BERSF=7,k=1,2,3AndSF=8,k=1,2,3} shows the BER performance of the conventional FSCSS system for $M=128$ and $M=256$ in an AWGN channel. A total of $10^9$ Monte-Carlo trials were used to generate the plots. A similar trend is observed in this figure, i.e., the lower the data rate of a scheme is, the better it performs in terms of BER. This trend is evident from the slight degradation in BER performance of FSCSS-IM when $|\mathcal{I}_m|$ increases from $1$ to $3$ for both values of $M$. Such performance degradation can be intuitively explained from the fact that in the conventional FSCSS scheme ($|\mathcal{I}_m|=1$), the index sets are non-overlapping, whereas they generally overlap in the proposed FSCSS-IM scheme ($|\mathcal{I}_m|>1$). This overlap results in slight degradation of the bit-error-rate performance.

\begin{figure}[t!]
\centering
\includegraphics[width=\columnwidth]{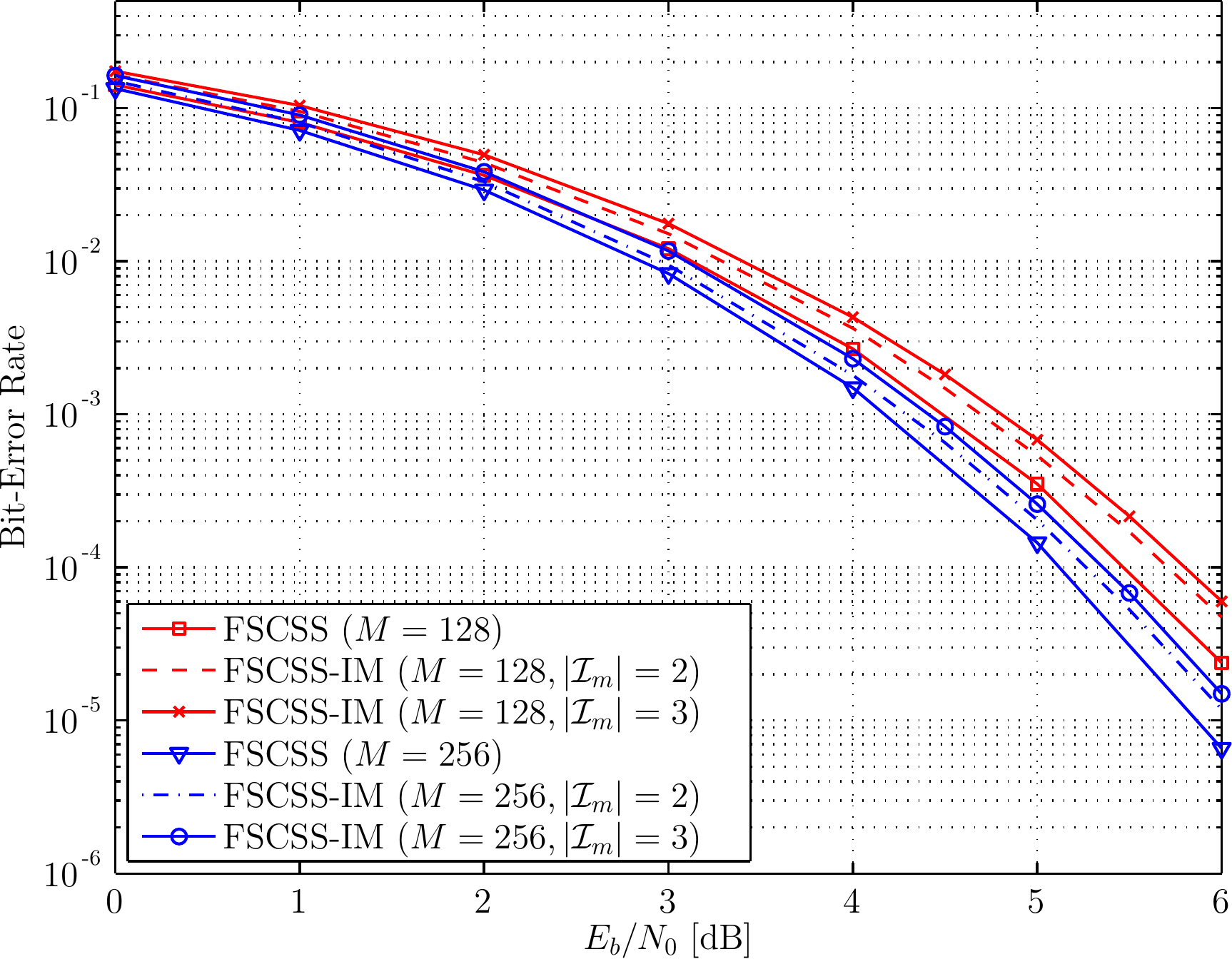}	
\caption{BER comparison of the conventional FSCSS and proposed FSCSS-IM systems for $M=128, 256$.}
\label{Fig:BERSF=7,k=1,2,3AndSF=8,k=1,2,3}
\end{figure}

Another interesting observation can be made from Fig. \ref{Fig:BERSF=7,k=1,2,3AndSF=8,k=1,2,3}. For each $M$, the bit-error rates for $|\mathcal{I}_m|=2,3$ are only slightly worse than that for $|\mathcal{I}_m|=1$, which corresponds to the conventional FSCSS modulation scheme. On the other hand, the maximum data rates achieved with $|\mathcal{I}_m|=2,3$ are \emph{significantly higher} than that with the conventional FSCSS modulation. For instance, the maximum data rate achieved with $|\mathcal{I}_m|=2$ and $|\mathcal{I}_m|=3$ for $M=128$ are $\Lambda/\log_2(M) = 12/7 \approx 1.7143$  and $\Lambda/\log_2(M) = 18/7 \approx 2.5714$ times the maximum data rate achieved by the conventional FSCSS modulation, respectively. So the proposed FSCSS-IM system can improve the achievable data rate significantly with only slight degradation in the bit-error rate. The data rate can further be improved by using more chirps for transmission and reception, i.e., by using $|\mathcal{I}_m| > 3$.

On a difference comparison, Fig. \ref{Fig:BERSF=7,k=1,2,3AndSF=8,k=1,2,3} actually shows that the proposed FSCSS-IM scheme can achieve a higher data rate than the conventional FSCSS modulation scheme while slightly improving the BER performance. Specifically, consider the BER performance of the conventional FSCSS modulation scheme for $M=128$ and the proposed FSCSS-IM scheme for $M=256$ and $|\mathcal{I}_m| = 3$. The figure depicts that the proposed FSCSS-IM scheme has slightly better BER performance than the conventional scheme. Note that the symbol duration of the conventional FSCSS modulation for $M=128$ is half of the symbol duration of the proposed FSCSS-IM scheme for $M=256$. Also, each transmitted symbol conveys a total of $\Lambda = 21$ bits in the proposed FSCSS-IM scheme. On the other hand, in the same symbol duration, the conventional FSCSS scheme can transmit a total of $2 \times 7 = 14$ bits. That is, the proposed FSCSS-IM scheme, while performing slightly better than the conventional FSCSS modulation scheme, can also achieve a data rate of $1.5$ times that of the conventional FSCSS modulation scheme.

In Fig. \ref{Fig:SERSF=7,k=1,2AndSF=8,k=1,2}, the symbol-error-rate (SER) performance of both the conventional FSCSS and proposed FSCSS-IM schemes are plotted against the channel SNR, $\bar{\gamma}$, in an AWGN channel. Observe that the lower the data rate of a scheme, the better the scheme is in terms of SER, and vice versa. In the figure, for example, $(M,|\mathcal{I}_m|)=(256,1)$ scheme has the lowest SER for a given channel SNR, while $(M,|\mathcal{I}_m|)=(128,2)$ has the highest SER. At the same time, the first scheme has the lowest data rate, whereas the latter has the highest data rate. The $(M,|\mathcal{I}_m|)=(128,1)$ and $(M,|\mathcal{I}_m|)=(256,2)$ schemes have the same data rate and have almost the same SER performance against the channel SNR. That is, the proposed FSCSS-IM scheme and conventional FSCSS modulation scheme have similar reception sensitivity for the same data rate.

\begin{figure}[t!]
\centering
\includegraphics[width=\columnwidth]{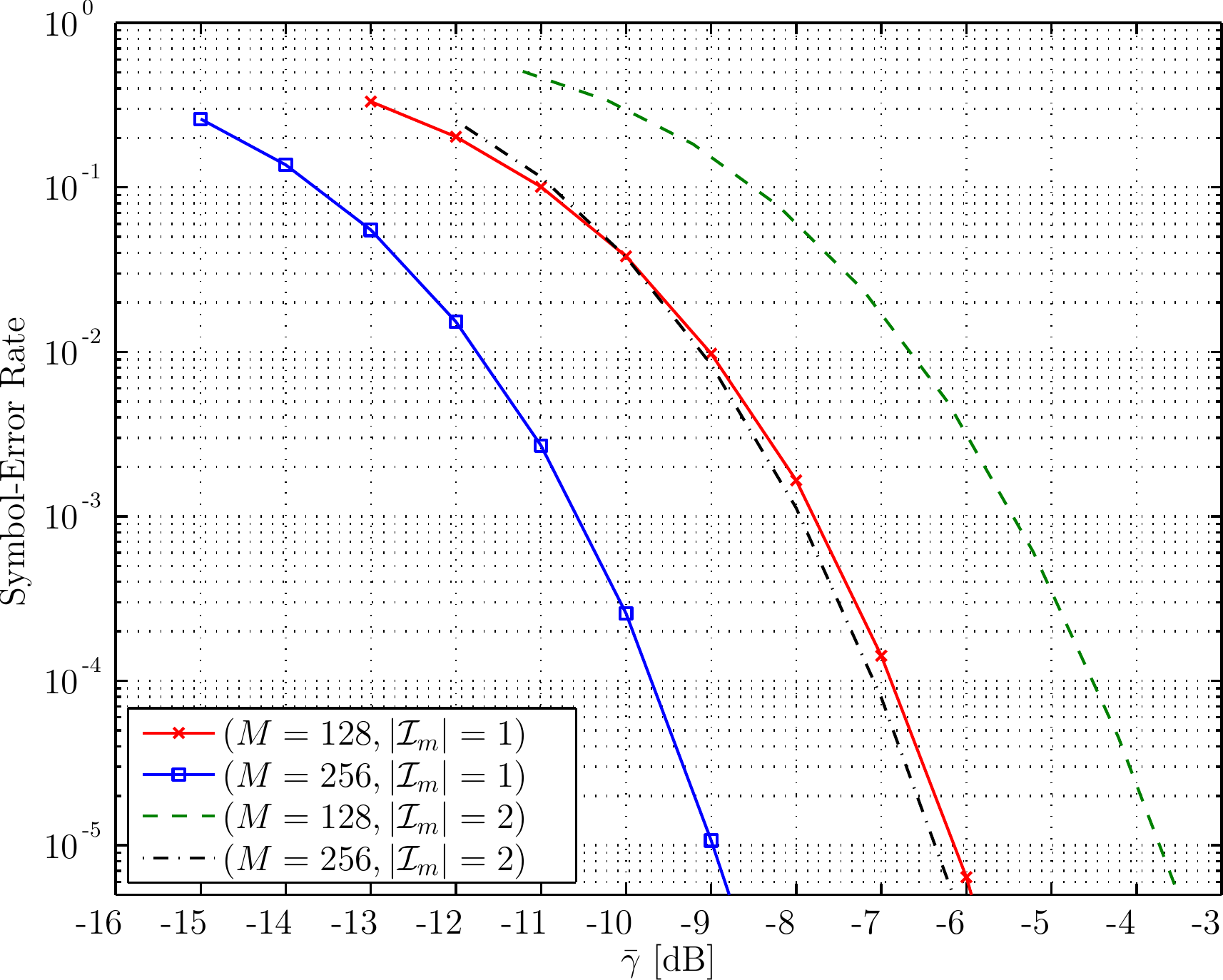}	
\caption{SER comparison of the conventional FSCSS modulation and proposed FSCSS-IM schemes for $M=128, 256$  and $|\mathcal{I}_m|=1, 2$.}
\label{Fig:SERSF=7,k=1,2AndSF=8,k=1,2}
\end{figure}

Next, Fig. \ref{Fig:SERSF=8,k=1,2,3SSKLoRaKMax} compares performance of the proposed FSCSS-IM scheme with the SSK-LoRa scheme introduced in \cite{SSKLoRa}. The plots are generated using a total of $10^9$ Monte-Carlo trials (i.e., $10^9$ FSCSS-IM or SSK-LoRa symbols are simulated for each SNR value). Fig. \ref{Fig:SERSF=8,k=1,2,3SSKLoRaKMax} depicts that the BER performance of SSK-LoRa modulation is better than the BERs of both the conventional FSCSS and proposed FSCSS-IM schemes. On the other hand, the proposed FSCSS-IM scheme performs slightly worse than the conventional FSCSS modulation scheme when the receiver of Algorithm \ref{Alg:ProposedReceiverPractical} is used. In terms of data rate, although the SSK-LoRa modulation scheme also achieves higher data rates than the conventional FSCSS modulation scheme, the data-rate improvements are marginal. In particular, for $M=256$, the data-rate improvement is only $12.5\%$. In contrast, the proposed FSCSS-IM scheme achieves significant improvements in the data rate while performing only slightly worse than the conventional FSCSS modulation scheme. The data-rate improvements for the proposed scheme, as given in Table \ref{Table:DataRateImprovements}, are $75 \%$ and $163 \%$ for $|\mathcal{I}_m|=2$, and $|\mathcal{I}_m|=3$, respectively. Finally, the figure also depicts the performance of the proposed FSCSS-IM scheme when using a non-coherent detection algorithm similar to that proposed in \cite{NoncoherentOFDMIM} for OFDM-IM and \cite{multiToneFSKOptimal} for multi-tone FSK modulation (labeled as ``K-Max'' in the figure's legend). In particular, in the K-max algorithm, the indexes of the largest $|\mathcal{I}_m|$ values of $|x_0^*[l]R[l]|$ (or equivalently, $|R[l]|$, since $|x_0^*[l]|=1$) constitute the estimated index set. Observe that our proposed detection scheme results in much better performance than the scheme of \cite{NoncoherentOFDMIM}.

\begin{figure}[t!]
\centering
\includegraphics[width=\columnwidth]{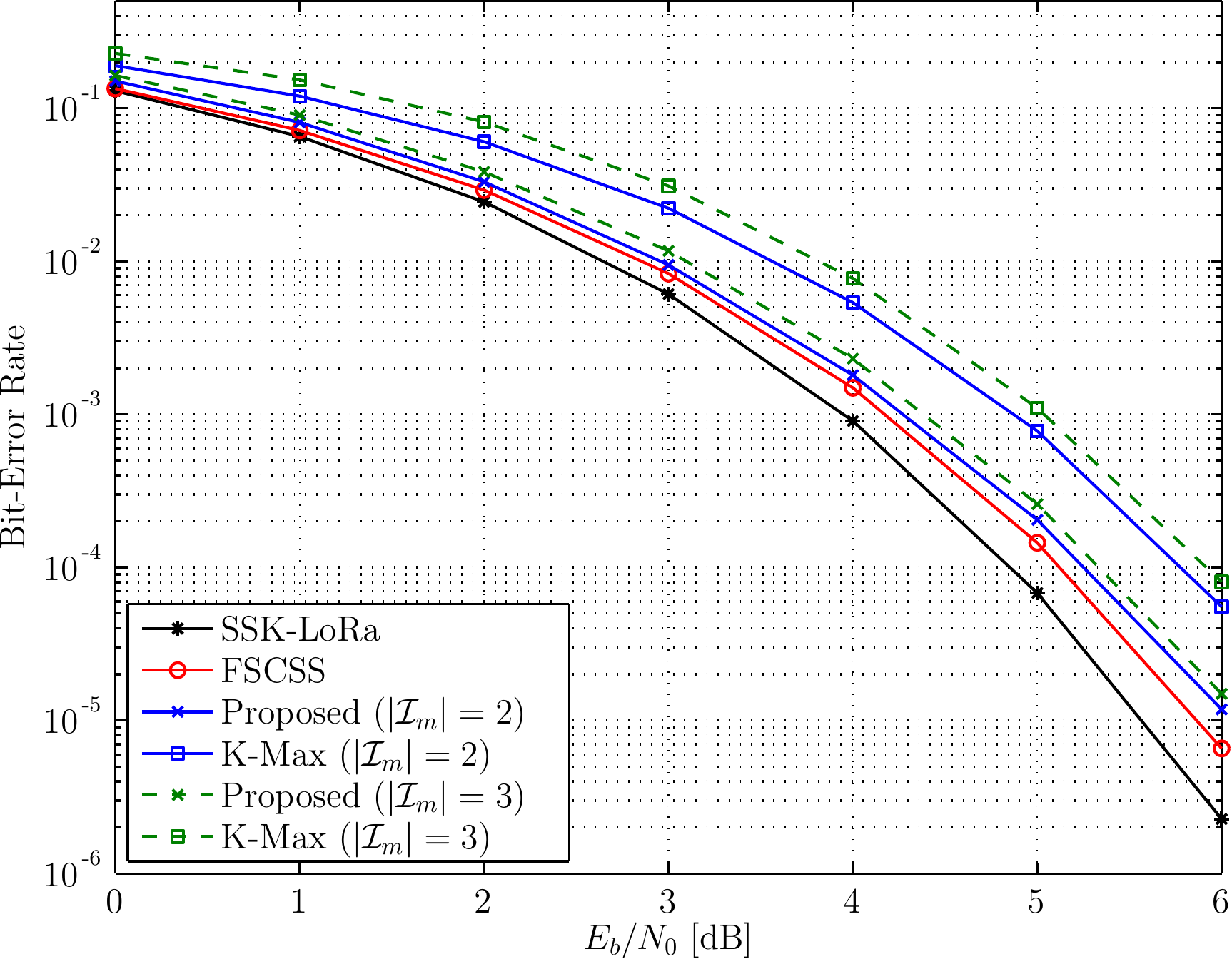}	
\caption{BER comparison of the FSCSS-based modulation schemes under different detection schemes for $M=256$.}
\label{Fig:SERSF=8,k=1,2,3SSKLoRaKMax}
\end{figure}

Finally, Fig. \ref{Fig:BERSF=7,k=2,Rayleigh} shows the performance of the proposed scheme over a Rayleigh fading channel. A total of $10^6$ Monte-Carlo trials were used for each SNR to generate the plot. The figure displays the performance of the optimal detection schemes along with the proposed suboptimal detection scheme. As expected, because the optimal coherent detector makes use of the channel-state information, it shows the best performance. The optimal non-coherent detection incurs a SNR degradation of about 0.8 dB as compared to the coherent detection, whereas the proposed suboptimal reception shows almost identical performance as that of the optimal non-coherent reception. Again, this comparison clearly demonstrates the efficiency and effectiveness of the proposed low-complexity non-coherent detection for the FSCSS-IM scheme.

\begin{figure}[t!]
\centering
\includegraphics[width=\columnwidth]{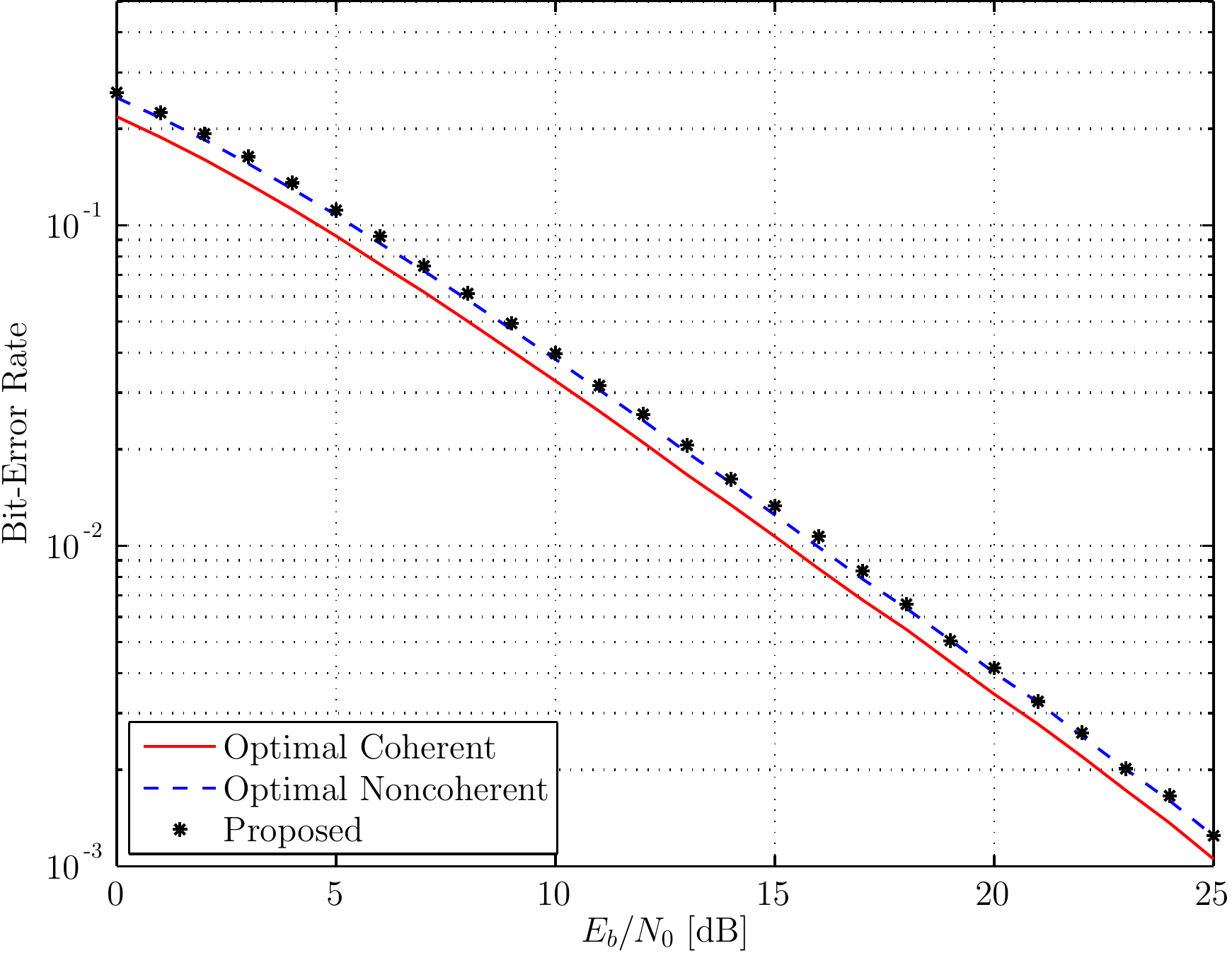}	 
\caption{BER performance of the proposed FSCSS-IM with different detection schemes for $M=128$, $|\mathcal{I}_m|=2$ over a Rayleigh fading channel.}
\label{Fig:BERSF=7,k=2,Rayleigh}
\end{figure}

\section{Conclusion}\label{sec:conclu}

In this paper, we introduced index modulation into the conventional FSCSS system to improve the achievable data rates for different system parameters. We have also developed optimal coherent and non-coherent detection algorithms for the proposed FSCSS-IM scheme under fading channels. Moreover, a suboptimal non-coherent detection algorithm was also proposed that eliminates memory requirement and significantly reduces the computational complexity of the optimal algorithm, while performing almost equally well. The proposed FSCSS-IM scheme is very flexible to design and can achieve much higher data rates than the conventional FSCSS system (under the same bandwidth) with only slight degradations in the bit-error rate performance.

Although the conventional FSCSS system uses orthogonal chirps, the concept of index modulation can also be applied for a set of non-orthogonal chirps. As discussed in the paper, using both up and down chirps, and even together with interleaved chirps can increase the highest achievable data rate of the conventional FSCSS modulation scheme. Moreover, applying the concept of index modulation on top of such an expanded set of (non-orthogonal) chirps can further improve the data rate. Developing optimal and low-complexity suboptimal receivers and performing theoretical analysis of the bit-error rate performance for such a novel scheme (as well as for the proposed FSCSS-IM for that matter) is an interesting topic and is left as a future research topic.

\balance

\section*{Acknowledgement}

This work is supported by the NSERC/Cisco Industrial Research Chair in Low-Power Wireless Access in Sensor Networks.

\bibliographystyle{IEEEtran}

\begin{thebibliography}{10}
\providecommand{\url}[1]{#1}
\csname url@samestyle\endcsname
\providecommand{\newblock}{\relax}
\providecommand{\bibinfo}[2]{#2}
\providecommand{\BIBentrySTDinterwordspacing}{\spaceskip=0pt\relax}
\providecommand{\BIBentryALTinterwordstretchfactor}{4}
\providecommand{\BIBentryALTinterwordspacing}{\spaceskip=\fontdimen2\font plus
\BIBentryALTinterwordstretchfactor\fontdimen3\font minus
  \fontdimen4\font\relax}
\providecommand{\BIBforeignlanguage}[2]{{%
\expandafter\ifx\csname l@#1\endcsname\relax
\typeout{** WARNING: IEEEtran.bst: No hyphenation pattern has been}%
\typeout{** loaded for the language `#1'. Using the pattern for}%
\typeout{** the default language instead.}%
\else
\language=\csname l@#1\endcsname
\fi
#2}}
\providecommand{\BIBdecl}{\relax}
\BIBdecl

\bibitem{IEEE802154a}
E.~{Karapistoli}, F.~{Pavlidou}, I.~{Gragopoulos}, and I.~{Tsetsinas}, ``An
  overview of the {IEEE} 802.15.4a standard,'' \emph{{IEEE} Commun. Mag.},
  vol.~48, no.~1, pp. 47--53, Jan. 2010.

\bibitem{LPWANTutorial}
U.~{Raza}, P.~{Kulkarni}, and M.~{Sooriyabandara}, ``Low power wide area
  networks: An overview,'' \emph{{IEEE} Commun. Surveys Tuts.}, vol.~19, no.~2,
  pp. 855--873, 2nd quarter 2017.

\bibitem{SemtechAN22}
SemTech, ``{AN1200.22: LoRa Modulation Basics},'' May 2015.

\bibitem{LoRaDesignGuide}
------, ``{AN1200.13: SX1272/3/6/7/8: LoRa Modem Designer’s Guide},'' Jul.
  2013.

\bibitem{FreqShiftChirpModLoRa}
L.~{Vangelista}, ``Frequency shift chirp modulation: The {LoRa} modulation,''
  \emph{{IEEE} Signal Process. Lett.}, vol.~24, no.~12, pp. 1818--1821, Dec.
  2017.

\bibitem{ChirpRadar}
J.~R. {Klauder}, A.~C. {Price}, S.~{Darlington}, and W.~J. {Albersheim}, ``The
  theory and design of chirp radars,'' \emph{The Bell Syst. Tech. J.}, vol.~39,
  no.~4, pp. 745--808, Jul. 1960.

\bibitem{FMChirpCommun}
D.~S. {Dayton}, ``{FM} ``chirp'' communications: Multiple access to dispersive
  channels,'' \emph{{IEEE} Trans. Electromagn. Compat.}, vol. EMC-10, no.~2,
  pp. 296--297, Jun. 1968.

\bibitem{ChirpModDigitalSignal}
A.~{Berni} and W.~{Gregg}, ``On the utility of chirp modulation for digital
  signaling,'' \emph{{IEEE} Trans. Commun.}, vol.~21, no.~6, pp. 748--751, Jun.
  1973.

\bibitem{DQPSKChirp}
A.~{Springer}, W.~{Gugler}, M.~{Huemer}, R.~{Koller}, and R.~{Weigel}, ``A
  wireless spread-spectrum communication system using {SAW} chirped delay
  lines,'' \emph{{IEEE} Trans. Microw. Theory Techn.}, vol.~49, no.~4, pp.
  754--760, Apr. 2001.

\bibitem{QAMChirp}
P.~{Zhang} and H.~{Liu}, ``An ultra-wide band system with chirp spread spectrum
  transmission technique,'' in \emph{6th Int. Conf. ITS Telecommun.}, Jun.
  2006, pp. 294--297.

\bibitem{CSSLongRange}
B.~{Reynders} and S.~{Pollin}, ``Chirp spread spectrum as a modulation
  technique for long range communication,'' in \emph{Symp. Commun. Veh.
  Technol.}, Nov. 2016, pp. 1--5.

\bibitem{InterleavedCSS}
T.~{Elshabrawy} and J.~{Robert}, ``Interleaved chirp spreading {LoRa}-based
  modulation,'' \emph{{IEEE} Internet Things J.}, vol.~6, no.~2, pp.
  3855--3863, Apr. 2019.

\bibitem{PSKLoRa}
R.~{Bomfin}, M.~{Chafii}, and G.~{Fettweis}, ``A novel modulation for {IoT}:
  {PSK-LoRa},'' in \emph{IEEE Veh. Technol. Conf.}, Apr. 2019, pp. 1--5.

\bibitem{EfficientDesignCSS}
T.~T. {Nguyen}, H.~H. {Nguyen}, R.~{Barton}, and P.~{Grossetete}, ``Efficient
  design of chirp spread spectrum modulation for low-power wide-area
  networks,'' \emph{IEEE Internet Things J.}, vol.~6, pp. 9503--9515, Dec.
  2019.

\bibitem{SSKLoRa}
M.~{Hanif} and H.~H. {Nguyen}, ``Slope-shift keying {LoRa}-based modulation,''
  \emph{{IEEE} Internet Things J.}, vol.~8, no.~1, pp. 211--221, Jan. 2021.

\bibitem{FlipLoRa}
Z.~{Xu}, S.~{Tong}, P.~{Xie}, and J.~{Wang}, ``{FlipLoRa}: Resolving collisions
  with up-down quasi-orthogonality,'' in \emph{IEEE Int. Conf. Sens. Commun.
  Netw. (SECON)}, 2020, pp. 1--9.

\bibitem{PermutationModulation}
D.~Slepian, ``Permutation modulation,'' \emph{Proc. IEEE}, vol.~53, no.~3, pp.
  228--236, Mar. 1965.

\bibitem{IMfor5GBook}
M.~Wen, X.~Cheng, and L.~Yang, \emph{Index Modulation for {5G} Wireless
  Communications}.\hskip 1em plus 0.5em minus 0.4em\relax Springer
  International Publishing, 2016.

\bibitem{FSKPermutationModulation}
H.~L. Schneider, ``Data transmission with {FSK} permutation modulation,''
  \emph{The Bell System Technical Journal}, vol.~47, no.~6, pp. 1131--1138,
  Jul. 1968.

\bibitem{OFDMIMBasar}
E.~{Ba\c{s}ar}, {\"{U}}.~{Ayg\"{o}l\"{u}}, E.~{Panay{\i}rc{\i}}, and H.~V.
  {Poor}, ``Orthogonal frequency division multiplexing with index modulation,''
  \emph{{IEEE} Trans. Signal Process.}, vol.~61, no.~22, pp. 5536--5549, Nov.
  2013.

\bibitem{EnhancedOFDMwithIM}
M.~{Wen}, B.~{Ye}, E.~{Basar}, Q.~{Li}, and F.~{Ji}, ``Enhanced orthogonal
  frequency division multiplexing with index modulation,'' \emph{{IEEE} Trans.
  Wireless Commun.}, vol.~16, no.~7, pp. 4786--4801, Jul. 2017.

\bibitem{MultipleModeOFDMwithIM}
M.~{Wen}, Q.~{Li}, E.~{Basar}, and W.~{Zhang}, ``Generalized multiple-mode
  {OFDM} with index modulation,'' \emph{{IEEE} Trans. Wireless Commun.},
  vol.~17, no.~10, pp. 6531--6543, Oct. 2018.

\bibitem{NoncoherentOFDMIM}
J.~{Choi}, ``Noncoherent {OFDM-IM} and its performance analysis,'' \emph{{IEEE}
  Trans. Wireless Commun.}, vol.~17, no.~1, pp. 352--360, Jan. 2018.

\bibitem{GeneralizedOFDMIM}
R.~{Fan}, Y.~J. {Yu}, and Y.~L. {Guan}, ``Generalization of orthogonal
  frequency division multiplexing with index modulation,'' \emph{{IEEE} Trans.
  Wireless Commun.}, vol.~14, no.~10, pp. 5350--5359, Oct. 2015.

\bibitem{PCSpreadSpectrum}
S.~{Sasaki}, J.~Zhu, and G.~{Marubayashi}, ``Performance of parallel
  combinatory spread spectrum multiple access communication systems,'' in
  \emph{IEEE Int. Symp. Personal, Indoor and Mobile Radio Commun.}, Sep. 1991,
  pp. 204--208.

\bibitem{SpatialModulationConf}
Y.~A. {Chau} and S.-H. {Yu}, ``Space modulation on wireless fading channels,''
  in \emph{IEEE Veh. Technol. Conf.}, vol.~3, Oct. 2001, pp. 1668--1671.

\bibitem{SpatialModulation}
R.~Y. {Mesleh}, H.~{Haas}, S.~{Sinanovic}, C.~W. {Ahn}, and S.~{Yun}, ``Spatial
  modulation,'' \emph{{IEEE} Trans. Veh. Technol.}, vol.~57, no.~4, pp.
  2228--2241, Jul. 2008.

\bibitem{GeneralizedSpatialMod}
A.~{Younis}, N.~{Serafimovski}, R.~{Mesleh}, and H.~{Haas}, ``Generalised
  spatial modulation,'' in \emph{Asilomar Conf. Signals, Syst. Comput.}, 2010,
  pp. 1498--1502.

\bibitem{GeneralisedSpatialModMultipleAntenna}
J.~{Wang}, S.~{Jia}, and J.~{Song}, ``Generalised spatial modulation system
  with multiple active transmit antennas and low complexity detection scheme,''
  \emph{{IEEE} Trans. Wireless Commun.}, vol.~11, no.~4, pp. 1605--1615, Apr.
  2012.

\bibitem{SpaceShifKeyingMIMO}
J.~{Jeganathan}, A.~{Ghrayeb}, L.~{Szczecinski}, and A.~{Ceron}, ``Space shift
  keying modulation for {MIMO} channels,'' \emph{{IEEE} Trans. Wireless
  Commun.}, vol.~8, no.~7, pp. 3692--3703, Jul. 2009.

\bibitem{GeneralizedSpaceShiftKeying}
J.~{Jeganathan}, A.~{Ghrayeb}, and L.~{Szczecinski}, ``Generalized space shift
  keying modulation for {MIMO} channels,'' in \emph{Int. Symp. Pers. Indoor
  Mobile Radio Commun.}, 2008, pp. 1--5.

\bibitem{MultiToneFSK}
C.~Luo, M.~Medard, and L.~Zheng, ``On approaching wideband capacity using
  multitone {FSK},'' \emph{{IEEE} J. Sel. Areas Commun.}, vol.~23, no.~9, pp.
  1830--1838, Sep. 2005.

\bibitem{PMSMandIMTutorial}
N.~{Ishikawa}, S.~{Sugiura}, and L.~{Hanzo}, ``50 years of permutation, spatial
  and index modulation: From classic {RF} to visible light communications and
  data storage,'' \emph{{IEEE} Commun. Surveys Tuts.}, vol.~20, no.~3, pp.
  1905--1938, 3rd Quarter 2018.

\bibitem{SurveySpatialMod}
M.~{Wen}, B.~{Zheng}, K.~J. {Kim}, M.~{Di Renzo}, T.~A. {Tsiftsis}, K.~{Chen},
  and N.~{Al-Dhahir}, ``A survey on spatial modulation in emerging wireless
  systems: Research progresses and applications,'' \emph{{IEEE} J. Sel. Areas
  Commun.}, vol.~37, no.~9, pp. 1949--1972, Sep. 2019.

\bibitem{IMTechniquesNextGen}
E.~{Basar}, M.~{Wen}, R.~{Mesleh}, M.~{Di Renzo}, Y.~{Xiao}, and H.~{Haas},
  ``Index modulation techniques for next-generation wireless networks,''
  \emph{IEEE Access}, vol.~5, pp. 16\,693--16\,746, Aug. 2017.

\bibitem{MultidimensionalIMfor5GandBeyond}
S.~D. {Tusha}, A.~{Tusha}, E.~{Basar}, and H.~{Arslan}, ``Multidimensional
  index modulation for {5G} and beyond wireless networks,'' \emph{Proc. IEEE},
  vol. 109, no.~2, pp. 170--199, Feb. 2021.

\bibitem{USpatent}
M.~Hanif and H.~H. Nguyen, ``Methods for improving flexibility and data rate of
  chirp spread spectrum systems in {LoRaWAN}.''

\bibitem{OverlapCSS}
T.~{Yoon}, S.~{Ahn}, S.~Y. {Kim}, and S.~{Yoon}, ``Performance analysis of an
  overlap-based {CSS} system,'' in \emph{Int. Conf. Adv. Commun. Technol.},
  vol.~1, Feb. 2008, pp. 424--426.

\bibitem{CodeIndexModulation}
G.~{Kaddoum}, M.~F.~A. {Ahmed}, and Y.~{Nijsure}, ``Code index modulation: A
  high data rate and energy efficient communication system,'' \emph{{IEEE}
  Commun. Lett.}, vol.~19, no.~2, pp. 175--178, Feb. 2015.

\bibitem{GeneralizedCodeIndexModulation}
G.~{Kaddoum}, Y.~{Nijsure}, and H.~{Tran}, ``Generalized code index modulation
  technique for high-data-rate communication systems,'' \emph{{IEEE} Trans.
  Veh. Technol.}, vol.~65, no.~9, pp. 7000--7009, Sep. 2016.

\bibitem{OFDMIMSpreadSpectrum}
Q.~{Li}, M.~{Wen}, E.~{Basar}, and F.~{Chen}, ``Index modulated {OFDM} spread
  spectrum,'' \emph{{IEEE} Trans. Wireless Commun.}, vol.~17, no.~4, pp.
  2360--2374, Apr. 2018.

\bibitem{LoRaModIoT}
M.~{Chiani} and A.~{Elzanaty}, ``On the {LoRa} modulation for {IoT}: Waveform
  properties and spectral analysis,'' \emph{IEEE Internet Things J.}, vol.~6,
  no.~5, pp. 8463--8470, Oct. 2019.

\bibitem{SemtechAN26}
SemTech, ``{AN1200.26: LoRa and FCC Part 15.247: Measurement Guidance},'' May
  2015.

\bibitem{SemtechSX127x}
------, ``{SX1276/77/78/79 - 137 MHz to 1020 MHz Low Power Long Range
  Transceiver},'' Aug. 2016.

\bibitem{GoldsmithWirelessCommun}
A.~Goldsmith, \emph{Wireless Communications}.\hskip 1em plus 0.5em minus
  0.4em\relax Cambridge, U.K.: Cambridge University Press, 2005.

\bibitem{multiToneFSKOptimal}
C.~Luo and M.~Medard, ``Performance of single-tone and two-tone frequency-shift
  keying for ultrawideband,'' in \emph{Asilomar Conf. Signals, Syst. Comput.},
  vol.~1, Nov. 2002, pp. 701--705.

\end{thebibliography}


\begin{IEEEbiography}
[{\includegraphics[width=1in,height=1.25in,clip,keepaspectratio]{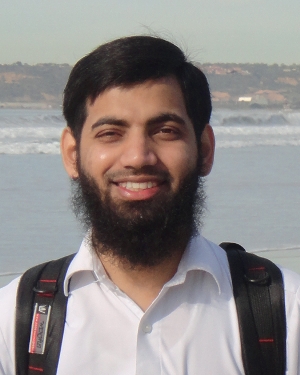}}]{Muhammad Hanif} received the Ph.D. degree in electrical engineering from the University of Victoria, Canada, in 2016. He was a Post-Doctoral Fellow with the University of Alberta, Canada, from 2016 to 2018, and with the University of Saskatchewan, Canada, from 2018 to 2019. He is currently an Assistant Professor at Thompson Rivers University, Canada. His research interests are in the general area of signal processing and wireless communication systems, including cognitive radio networks, massive MIMO systems, polar codes, and machine-to-machine communications.
\end{IEEEbiography}

\begin{IEEEbiography}
[{\includegraphics[width=1in,height=1.25in,clip,keepaspectratio]{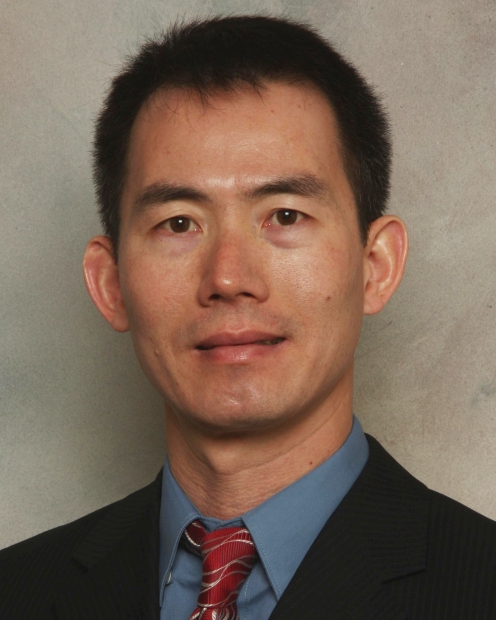}}]{Ha H. Nguyen} received the B.Eng. degree from Hanoi University of Technology (HUT), Hanoi, Vietnam, in 1995, the M.Eng. degree from the Asian Institute of Technology (AIT), Bangkok, Thailand, in 1997, and the Ph.D. degree from the University of Manitoba, Winnipeg, MB, Canada, in 2001, all in electrical engineering. He joined the Department of Electrical and Computer Engineering, University of Saskatchewan, Saskatoon, SK, Canada, in 2001, and became a full Professor in 2007. He currently holds the position of \emph{NSERC/Cisco Industrial Research Chair in Low-Power Wireless Access for Sensor Networks}. His research interests fall into broad areas of Communication Theory, Wireless Communications, and Statistical Signal Processing. Dr. Nguyen was an Associate Editor for the \emph{IEEE Transactions on Wireless Communications} and \emph{IEEE Wireless Communications Letters} during 2007-2011 and 2011-2016, respectively. He currently serves as an Associate Editor for the \emph{IEEE Transactions on Vehicular Technology}. He served as a Technical Program Chair for numerous IEEE events. He is a coauthor, with Ed Shwedyk, of the textbook "A First Course in Digital Communications" (published by Cambridge University Press). Dr. Nguyen is a Fellow of the Engineering Institute of Canada (EIC) and a Registered Member of the Association of Professional Engineers and Geoscientists of Saskatchewan (APEGS).
\end{IEEEbiography}

\end{document}